\documentclass[11pt]{article}
\usepackage{amsmath,amsthm,amssymb,cite,enumerate}
\usepackage[a4paper,left=0.5in,right=1in]{geometry}
\usepackage[usenames]{color}
\usepackage{graphicx}
\usepackage{epsfig}
\usepackage{dcolumn}
\usepackage{bm}
\usepackage{authblk} 
\usepackage{soul} 

\usepackage{t1enc}	

\begin{document}

\def \d {{\rm d}}

\def \bm {\mbox{\boldmath{$m$}}}
\def \bmt {\mbox{\boldmath{$\tilde m$}}}

\def \bF {\mbox{\boldmath{$F$}}}
\def \bA {\mbox{\boldmath{$A$}}}
\def \cF {\mbox{\boldmath{$\cal F$}}}
\def \bH {\mbox{\boldmath{$H$}}}
\def \bC {\mbox{\boldmath{$C$}}}
\def \bSS {\mbox{\boldmath{$S$}}}
\def \bS {\mbox{\boldmath{${\cal S}$}}}
\def \bV {\mbox{\boldmath{$V$}}}
\def \bff {\mbox{\boldmath{$f$}}}
\def \bT {\mbox{\boldmath{$T$}}}
\def \bk {\mbox{\boldmath{$k$}}}
\def \bl {\mbox{\boldmath{$\ell$}}}
\def \bn {\mbox{\boldmath{$n$}}}
\def \bbm {\mbox{\boldmath{$m$}}}
\def \tbbm {\mbox{\boldmath{$\bar m$}}}
\def \bet {\mbox{\boldmath{$\eta$}}}
\def \H {{\cal H}}

\def \T {\bigtriangleup}
\newcommand{\msub}[2]{m^{(#1)}_{#2}}
\newcommand{\msup}[2]{m_{(#1)}^{#2}}

\newcommand{\be}{\begin{equation}}
\newcommand{\ee}{\end{equation}}

\newcommand{\beqn}{\begin{eqnarray}}
\newcommand{\eeqn}{\end{eqnarray}}
\newcommand{\AdS}{anti--de~Sitter }
\newcommand{\AAdS}{\mbox{(anti--)}de~Sitter }
\newcommand{\AAN}{\mbox{(anti--)}Nariai }
\newcommand{\AS}{Aichelburg-Sexl }
\newcommand{\pa}{\partial}
\newcommand{\pp}{{\it pp\,}-}
\newcommand{\ba}{\begin{array}}
\newcommand{\ea}{\end{array}}

\newcommand*\bR{\ensuremath{\boldsymbol{R}}}

\newcommand*\BF{\ensuremath{\boldsymbol{F}}}
\newcommand*\BR{\ensuremath{\boldsymbol{R}}}
\newcommand*\BS{\ensuremath{\boldsymbol{S}}}
\newcommand*\BC{\ensuremath{\boldsymbol{C}}}
\newcommand*\bg{\ensuremath{\boldsymbol{g}}}
\newcommand*\bE{\ensuremath{\boldsymbol{E}}}

\newcommand*\bh{\ensuremath{\boldsymbol{h}}}
\newcommand*\bZ{\ensuremath{\boldsymbol{Z}}}

\def \bo {\mbox{\boldmath{$\omega$}}}
\def \bot {\mbox{\boldmath{$\tilde\omega$}}}
\def \bE {\mbox{\boldmath{$e$}}}
\def \bEt {\mbox{\boldmath{$\tilde e$}}}
\def \bG {\mbox{\boldmath{$\Gamma$}}}
\def \bGt {\mbox{\boldmath{$\tilde\Gamma$}}}
\def \bTt {\mbox{\boldmath{$\tilde\Theta$}}}

\newcommand{\M}[3] {{\stackrel{#1}{M}}_{{#2}{#3}}}
\newcommand{\m}[3] {{\stackrel{\hspace{.3cm}#1}{m}}_{\!{#2}{#3}}\,}

\newcommand{\tr}{\textcolor{red}}
\newcommand{\tb}{\textcolor{blue}}
\newcommand{\tg}{\textcolor{green}}

\newcommand{\thorn}{\mathop{\hbox{\rm \th}}\nolimits}

\def\a{\alpha}
\def\g{\gamma}
\def\de{\delta}

\def\E{{\cal E}}
\def\B{{\cal B}}
\def\R{{\cal R}}
\def\F{{\cal F}}
\def\L{{\cal L}}

\def\e{e}
\def\bb{b}

\newtheorem{theorem}{Theorem}[section] 
\newtheorem{cor}[theorem]{Corollary} 
\newtheorem{lemma}[theorem]{Lemma} 
\newtheorem{proposition}[theorem]{Proposition}
\newtheorem{definition}[theorem]{Definition}
\newtheorem{remark}[theorem]{Remark}

\title{Charging Kerr-Schild spacetimes in higher dimensions}

\author[1]{Marcello Ortaggio\thanks{ortaggio(at)math(dot)cas(dot)cz}}

\author[1,2]{Aravindhan Srinivasan\thanks{srinivasan(at)math(dot)cas(dot)cz}}

\affil[1]{Institute of Mathematics, Czech Academy of Sciences, \newline \v Zitn\' a 25, 115 67 Prague 1, Czech Republic}

\affil[2]{Institute of Theoretical Physics, Faculty of Mathematics and Physics, \newline
 Charles University, V Hole\v{s}ovi\v{c}k\'{a}ch 2, 180 00 Prague 8, Czech Republic}

\maketitle

\abstract{ 
We study higher dimensional charged Kerr-Schild (KS) spacetimes that can be constructed by a KS transformation of a vacuum solution with an arbitrary cosmological constant, and for which the vector potential is aligned with the KS vector $\bk$. Focusing on the case of an {\em expanding} $\bk$, we first characterize the presence of shear as an obstruction to non-null fields (thereby extending an early no-go result of Myers and Perry). We next obtain the complete family of shearfree solutions. In the twistfree case, they coincide with charged Schwarzschild-Tangherlini-like black holes. Solutions with a twisting $\bk$ consist of a four-parameter family of higher dimensional charged Taub-NUT metrics with a base space of constant holomorphic sectional curvature.
In passing, we identify the configurations for which the test-field limit gives rise to instances of the KS double copy. Finally, it is shown that null fields define a branch of twistfree but shearing solutions, exemplified by the product of a Vaidya-like radiating spacetime with an extra dimension. 
}

\vspace{.2cm}
\noindent

%

\tableofcontents

\section{Introduction}

\label{intro}

\subsection{Background}

\label{subsec_backgr}

The celebrated vacuum black hole metric of Kerr \cite{Kerr63} can be written as
\be
	\bg=\bet-2H\bk\otimes\bk ,  
	\label{KS}
\ee
where the ``background'' metric 
\be
 \bet=-\d u^2+2\d r(\d u+a\sin^2\theta\d\phi)+(r^2+a^2\cos^2\theta)\d\theta^2+(r^2+a^2)\sin^2\theta\d\phi^2			, 
 \label{eta_Kerr}
\ee
is flat, the covector field 
\be
 \bk=\d u+a\sin^2\theta\d\phi , 
 \label{k_Kerr}
\ee
is null (with respect to both metrics $\bet$ and $\bg$), and the scalar function $H$ is given by
\be
 2H=2H_{\mbox{\tiny Kerr}}\equiv-\frac{2mr}{r^2+a^2\cos^2\theta}	. 
 \label{H_Kerr}
\ee

It is remarkable that the charged Kerr-Newman solution \cite{Newmanetal65} can be described by exactly the same metric~\eqref{KS}--\eqref{k_Kerr} provided one takes a vector potential of the form 
\be
 \bA=-\frac{er}{r^2+a^2\cos^2\theta} \bk	, 
 \label{A_KN}
\ee
and simply replaces~\eqref{H_Kerr} in \eqref{KS} by the new function\footnote{In order to match the standard four-dimensional conventions \cite{Stephanibook,GriPodbook}, both in~\eqref{H_KN} and in appendix~\ref{app_KNAdS} we set the gravitational constant $\kappa=2$ (cf.~\eqref{Einst}). However, in the rest of the paper we find it more convenient to leave it unspecified.}
\be
 2H=2H_{\mbox{\tiny KN}}\equiv-\frac{2mr-e^2}{r^2+a^2\cos^2\theta}	. 
 \label{H_KN}
\ee
One can also easily obtain the Kerr-Newman-(A)dS solution from the Kerr-(A)dS metric with a similar method (where $\bet$ is now the (A)dS metric, cf. appendix~\ref{app_KNAdS} and refs. therein).

More generally, line-element~\eqref{KS} with $\bet$ flat and $\bk$ null (but not necessarily of the form \eqref{eta_Kerr}, \eqref{k_Kerr}, and with $H$ a priori unspecified) defines the Kerr-Schild (KS) class of spacetimes \cite{KerSch652} (see also the earlier \cite{Trautman62}), which clearly includes both the Kerr and the Kerr-Newman metrics as particular members. 
 It readily follows from the results of \cite{DebKerSch69} (see also \cite{Stephanibook}) that, similarly as in the Kerr case, all diverging vacuum KS metrics can be charged by simply taking 
\be
 \bA=\alpha\bk ,
 \label{A_KS}
\ee 
for a suitable spacetime function $\alpha$, and accordingly redefining $H$
\be
 H=H_{\mbox{\tiny vac}}\rightarrow H_{\mbox{\tiny elec}} ,
 \label{H_redef}
\ee
such as to keep into account the backreaction of the electromagnetic field in the Einstein equation. The form of the metric~\eqref{KS} (in particular $\bk$) is otherwise unchanged in the charging process.\footnote{This can be also understood as a particular case of a KS transformation \cite{Edelen66_I,Xanthopoulos83,ColHilSen01}. However, there also exist electrovac KS solutions for which $\bA$ is not of the form~\eqref{A_KS} \cite{DebKerSch69,Stephanibook}.}

In view of such a simple relation connecting vacuum KS solutions to their charged counterparts, it is desirable to clarify whether and to what extent a similar method can be used to produce electrovac solutions also in spacetime dimensions other than four. This is particularly relevant in the context of black holes, given that the higher-dimensional analogues of the Kerr black hole, i.e., the Myers-Perry spacetimes, are precisely of the KS form \cite{MyePer86} (see also \cite{Chakrabarti86} in eight dimensions),\footnote{That this is not the case for black rings \cite{EmpRea02prl} has been proven in \cite{PraPra05,OrtPraPra09}.} and that a similar property holds also in the presence of a cosmological constant $\Lambda$ \cite{HawHunTay99,Gibbonsetal05}.\footnote{In the latter case, it is understood that in~\eqref{KS} $\bet$ is a spacetime of constant non-zero curvature, such that \eqref{KS} defines the KS-(A)dS class. For the sake of brevity, however, throughout this paper we shall simply call any $\bg$ of the form~\eqref{KS} a KS spacetime, provided $\bet$ is of constant (positive, negative or zero) curvature. In order to disregard the trivial (vacuum) case $\bg=\bet$, it will also be understood that $H\neq0$.}

An early attempt at obtaining charged rotating black holes within the KS class was already performed in \cite{MyePer86}.\footnote{When the angular momentum vanishes, a static charged black hole (with an arbitrary $\Lambda$) has been known for some time \cite{Tangherlini63}, and it is easy to see it indeed admits a KS representation (e.g. using Robinson-Trautman coordinates \cite{PodOrt06}, cf. also section~\ref{subsec_RT}). Five-dimensional charged black holes with non-zero angular momentum exist \cite{KunNavPer05,KunNavVie06,Frobetal22} which do {\em not} belong to the KS class \cite{Frobetal22}.} It was concluded there that for metrics with precisely one non-zero spin, the Maxwell and the Einstein equations become incompatible if the KS ansatz is assumed. However, one might hope that such a no-go result could be circumvented in a more general context, e.g. by adding spins or a cosmological constant to the seed black hole. There are indeed several reasons why an ansatz as~\eqref{KS} with \eqref{A_KS}, \eqref{H_redef} seems a promising one in order to construct a charged solution from a vacuum one, and which motivate its further exploration in an arbitrary number of dimensions $n$. The most important ones, which will be useful throughout the paper, are summarized below.

\begin{enumerate}

	\item\label{k_geod} If $\bg$ in~\eqref{KS} is an Einstein spacetime, then the null congruence defined by $\bk$ must be {\em geodesic}. More generally, geodesicity is equivalent to the much milder condition $R_{ab}k^ak^b=0$ \cite{OrtPraPra09,MalPra11}. The geodesic property holds simultaneously in any geometry~\eqref{KS} (i.e., regardless of the choice of $H$, including the background with $H=0$). It plays a crucial role in what follows and will be understood from now on. Similarly, also the optical matrix of $\bk$ (cf. section~\ref{subsec_opt_matr}) does not depend on the function $H$, and must obey the {\em optical constraint} \cite{OrtPraPra09,MalPra11} (this is defined in~\eqref{OC} below, cf. \cite{OrtPraPra09,OrtPraPra10,MalPra11,OrtPraPra13,OrtPraPra13rev}, and becomes trivial if $\bk$ is non-expanding).

	\item\label{k_Riem} For any $H$, the Riemann tensor of the geometry $\bg$ is of type II (or more special), aligned with $\bk$  \cite{OrtPraPra09,MalPra11} in the classification of \cite{Milsonetal05} (cf. also \cite{OrtPraPra13rev}), which thus also defines a multiple Weyl aligned null direction (mWAND) \cite{Coleyetal04}. Imposing the Einstein equation, this implies, in particular, that the energy-momentum tensor of any geometry $ds^2$ must satisfy $T_{ab}k^ak^b=0=k^aT_{a[b}k_{c]}$ (i.e., it is also of type II or more special). This condition is compatible, in particular, with a Maxwell field of the form $\bF=\d\bA$ with \eqref{A_KS}, which is automatically aligned with $\bk$ (i.e., $F^a_{\phantom{a}b}k^b\propto k^a$ or, equivalently, $k^aF_{a[b}k_{c]}=0$; cf. also \cite{HerOrtWyl13}).

	\item The mixed Ricci components $R^a_{\phantom{a}b}$ are linear in the function $H$ and its derivatives \cite{GurGur75,DerGur86}. The same is true for the frame Ricci components \cite{OrtPraPra09,MalPra11} (cf. appendix~\ref{app_Riem}).

	\item\label{k_Maxw} One can easily see that $\sqrt{-g}=\sqrt{-\eta}$ and (with \eqref{A_KS}) $g^{ac}g^{bd}F_{cd}=\eta^{ac}\eta^{bd}F_{cd}$, therefore the Maxwell equation for $\bF$, i.e., $(\sqrt{-g}F^{ab})_{,b}=0$, is independent of $H$ -- this means that one can solve it in the background $\bet$, thereby obtaining a test field solution  $\bF$ valid for any $H$. The specific form of $H$ will be subsequently determined by solving the Einstein equation~\eqref{Einst}, \eqref{T}, thus arriving at a full Einstein-Maxwell solution which keeps into account the backreaction. In the case $\Lambda=0$ this was already observed in \cite{MyePer86}.

\end{enumerate}

The above results were already known for the special case $n=4$ and are scattered in various papers \cite{KerSch652,Thompson66,Edelen66_I,DebKerSch69,Dozmorov70,GurGur75,Xanthopoulos78,Taub81,Xanthopoulos83} (see also \cite{Stephanibook}).

It is also worth mentioning that the ansatz \eqref{KS} with \eqref{A_KS} is further motivated from the viewpoint of the KS double copy \cite{MonOCoWhi14} (further comments will be given in sections~\ref{subsubsec_shearfree} and \ref{subsec_RT}).

\subsection{Summary of results}

The present contribution aims at classifying all higher dimensional electrovac KS spacetimes that can be obtained by charging a vacuum KS spacetime via the ansatz~\eqref{KS}, \eqref{A_KS} with the redefinition~\eqref{H_redef}, where $\bk$ is null and {\em expanding} (cf. section~\ref{subsec_notat}), and $\bet$ of constant curvature. After preliminarily defining certain key quantities and setting up the field equations for the KS ansatz in section~\ref{sec_equations}, our main results can be summarized as follows (see also table~\ref{table_summary}).

\begin{itemize}

	\item If $\bk$ is twisting (section~\ref{sec_twisting}), then it must be {\em shearfree} and $\bF=\d\bA$ is necessarily {\em non-null}. This is possible only in {\em even} dimensions and, as it turns out, the only such solution is given by the special charged Taub-NUT solution~\eqref{shearfree_g}--\eqref{DF2}. This is specified by four independent parameters $\mu$, $\ell$, $\alpha_0$ and $\beta_0$  (roughly corresponding to mass, NUT, electric charge and asymptotic magnetic field strength), in addition to an arbitrary cosmological constant.\footnote{Vacuum spacetimes with multiple NUT parameters \cite{ManSte04} cannot be charged with the procedure employed in this paper precisely because they are (twisting and) shearing \cite{OrtPraPra13,Ortaggio17,Taghavi-Chabert21,Alekseevskyetal21} (although some of them belong to the KS class \cite{Ortaggio17}). For the same reason, the charged version of the odd-dimensional solutions of \cite{ManSte04} is also ruled out (cf.~\cite{OrtPraPra07}).} The base space metric $\bh$ must be K\"ahler-Einstein and of constant holomorphic sectional curvature (cf.~\eqref{NUT_KS_cond2}). We observe that the no-go result of \cite{MyePer86} was obtained for a KS ansatz based on a shearing $\bk$, which explains why there is no contradiction with the KS solutions found here.
	
	\item If $\bk$ is twistfree (section~\ref{sec_twistfree}) there are two possibilities. (i) For zero shear, one is left with a subclass of the Robinson-Trautman electrovac solutions \cite{OrtPodZof08}, for which $\bF$ is again {\em non-null}. It eventually reduces to the solution~\eqref{RT_g}, \eqref{RT_F}, which describes electrically charged Schwarzschild-Tangherlini black holes \cite{Tangherlini63,GibWil87} (see also \cite{KodIsh04}), characterized by mass and charge parameters $\mu$ and $\alpha_0$, and with a base space of constant curvature. (ii) For non-zero shear, $\bF$ is instead {\em null}. Solutions in this branch, which include products of a Vaidya-like radiating spacetime with flat extra dimensions, will be studied in more detail elsewhere.

\end{itemize}

\begin{table}[t]
\begin{center}
\begin{tabular}{|r|r|c|c|c|} 
  \hline 
 $\omega$ & \quad $\sigma$ \quad  & \quad $\bF$ \quad & { solution} & $n$  \\ \hline\hline   
  $\neq0$ & $0$  & non-null  & \eqref{shearfree_g}--\eqref{DF2} & even \\ \hline 
	$0$			& $0$ & non-null & \eqref{RT_g}, \eqref{RT_F} & any \\ \hline 
	$0$			& $\neq0$ & null &  \eqref{string_g}, \eqref{string_F} & any \\ \hline  						
\end{tabular} 
\end{center}
		\caption{{\footnotesize Summary of the $n>4$ charged KS solutions obtained with the ansatz~\eqref{KS}, \eqref{A_KS}, \eqref{H_redef}, where $\bk$ is assumed to be expanding ($\theta\neq0$). The optical scalars are defined in~\eqref{scalars}. Note that the solution~\eqref{string_g}, \eqref{string_F} only represents a particular five-dimensional example for the null field branch~$\omega=0\neq\sigma$ (which will deserve a separate study elsewhere).}}
 \label{table_summary}
\end{table}

The appendices contain several auxiliary results. Appendix~\ref{app_KNAdS} reviews the KS form of the four-dimensional Kerr-Newman-(A)dS solution and its ``topological'' counterparts  \cite{KleMorVan98,CalKle99}. Appendix~\ref{app_Riem} presents the connection (Ricci rotation coefficients) and the Riemann and Ricci tensors \cite{OrtPraPra09,MalPra11} in an arbitrary adapted frame for KS spacetimes with a geodesic $\bk$ in $n$ dimensions. Appendix~\ref{app_NUT} summarizes basic properties of the higher-dimensional Taub-NUT metrics of \cite{Berard82,BaiBat85,PagPop87}, and reviews the integration of the corresponding Einstein equation in vacuum. A few new observations are also added, in particular concerning the overlap with the KS class. Finally, appendix~\ref{app_integration} contains the details of the integration of the Einstein-Maxwell equations for the case of a shearfree twisting KS vector field $\bk$, relevant to the branch of solutions of section~\ref{subsubsec_shearfree}.

\section{Preliminaries and field equations}

\label{sec_equations}

\subsection{Notation}

\label{subsec_notat}

We will consider the Einstein-Maxwell equations in $n$ spacetime dimensions in the form
\beqn
 & & G_{ab}+\Lambda g_{ab}=\kappa T_{ab} , \label{Einst} \\ 
 & & \nabla_b{F}^{ab}=0 , \label{Maxw}
\eeqn
where $G_{ab}$ is the Einstein tensor and 
\be 
	T_{ab}=F_{ac}F_b^{\phantom{b}c}-\frac{1}{4}g_{ab}F_{cd}F^{cd} .
	\label{T}
\ee 
It will be also convenient to define a rescaled cosmological constant as
\be
 \lambda\equiv\frac{2\Lambda}{(n-1)(n-2)} .
 \label{lambda}
\ee

We will use a frame adapted to the KS ansatz~\eqref{KS}, i.e., a set of $n$ vectors $\bm_{(a)} $ which consists of two null vectors $\bk\equiv\bm_{(0)}$,  $\bn\equiv\bm_{(1)}$ and $n-2$ orthonormal spacelike vectors $\bm_{(i)} $, with $a, b\ldots=0,\ldots,n-1$ while $i, j  \ldots=2,\ldots,n-1$ \cite{Coleyetal04,OrtPraPra13rev}).\footnote{With a slight abuse of notation, we will use the symbol $\bk$ to denote both the vector field $k^a\pa_a$ and the corresponding covector $k_a\d x^a$ (where $k_a=g_{ab}k^b$). In all cases, it will be clear from the context what is the object under consideration.} The Ricci rotation coefficients $L_{ab}$, $N_{ab}$ and $\M{i}{a}{b}$ are defined by (the following quantities are meant as projections onto the basis vectors) \cite{Pravdaetal04}
\be
 L_{ab}=k_{a;b} , \qquad N_{ab}=n_{a;b}  , \qquad \M{i}{a}{b}=m^{(i)}_{a;b} ,
 \label{Ricci_rot}
\ee
and satisfy the identities $L_{0a}=N_{1a}=N_{0a}+L_{1a}=\M{i}{0}{a} + L_{ia} = \M{i}{1}{a}+N_{ia}=\M{i}{j}{a}+\M{j}{i}{a}=0$. 

Since by construction $\bk$ is geodesic (cf. point~\ref{k_geod}. in section~\ref{subsec_backgr}), with no loss of generality we will also assume it is affinely parametrized, and define an affine parameter $r$ such that
\be
 k^a\pa_a=\pa_r .
 \label{affine}
\ee
By also using a frame parallelly transported along $\bk$ \cite{OrtPraPra07}, one thus has
\be
 L_{i0}=L_{10}=\M{i}{j}{0}=N_{i0}=0 . 
\label{pt}
\ee
Furthermore, it also follows that the rank of the {\em optical matrix}
\be
  L_{ij}=k_{a;b}m_{(i)}^am_{(j)}^b ,
	\label{L_def}
\ee
is a frame-independent property  (cf. section~2.2 of \cite{OrtPraPra07}). 

The symmetric and antisymmetric parts of the optical matrix 
\be
  S_{ij}\equiv L_{(ij)}=\sigma_{ij}+\theta\delta_{ij} , \qquad A_{ij}\equiv L_{[ij]} ,
	\label{S_A}
\ee
where $\sigma_{ii}=0$, can be used to define the expansion, shear and twist scalars, namely \cite{Pravdaetal04,OrtPraPra07}
\be
	\theta\equiv\textstyle{\frac{1}{n-2}}S_{ii} , \qquad \sigma^2\equiv\sigma_{ij}\sigma_{ij}, \qquad \omega^2\equiv A_{ij}A_{ij} .
	\label{scalars}
\ee

Covariant derivatives along the frame vectors are denoted as
\be
	D \equiv k^a \nabla_a, \qquad \T\equiv n^a \nabla_a, \qquad \delta_i \equiv m_{(i)}^{a} \nabla_a . 
 \label{covder}
\ee

Throughout the paper we will consider only solutions for which $\bk$ is {\em expanding}, i.e., from now on we assume $\theta\neq0$. In addition, $\bk$ will define a null direction doubly aligned with the Riemann tensor (see point~\ref{k_Riem}. in section~\ref{subsec_backgr}), i.e.,
\be
 R_{0i0j}=0 , \qquad R_{0ijk}=0=R_{010i} .
 \label {typeII}
\ee

Upon using~\eqref{pt} and \eqref{typeII}, a subset of the components of the Ricci identity \cite{OrtPraPra07} take the form
\beqn
  & & DL_{1i}=-L_{1j}L_{ji} , \qquad  D L_{i1}=-L_{ij}L_{j1} , \label{11e} \\
	& & D L_{ij}=- L_{ik} L_{kj} , \label{sachs} \\
	& & \delta_{[j|} L_{i|k]} =  L_{1[j|} L_{i|k]} + L_{i1} L_{[jk]}   +  L_{il} \M{l}{[j}{k]}+ L_{l[j|}  \M{l}{i|}{k]} ,  \label{11k}
\eeqn
while the commutators of the derivative operators \cite{Coleyetal04vsi} reduce to
\beqn
 & & \T D - D \T=L_{11} D + L_{i1} \delta_i , \label{comm_DelD} \\
 & & \delta_i D-D\delta_i=L_{1i}D+L_{ji}\delta_j , \label{comm_dD} \\
 & & \delta_i \T - \T \delta_i  = N_{i1} D + (L_{i1}-L_{1i}) \T + (N_{ji}+\M{j}{i}{1}) \delta_j , \label{comm_dDel} \\
 & & \delta_{[i}\delta_{j]}=N_{[ij]}D+L_{[ij]}\Delta+\M{k}{[i}{j]}\delta_k . \label{comm_dd}
\eeqn

\subsection{Optical matrix}

\label{subsec_opt_matr}

Throughout the paper a frame as described in section~\ref{subsec_notat} will be employed. The optical matrix~\eqref{L_def} of a vacuum KS seed must obey the optical constraint \cite{OrtPraPra09,MalPra11}
\be
	L_{ik}L_{jk}=\frac{L_{lk}L_{lk}}{(n-2)\theta}S_{ij} .
	\label{OC}
\ee
Together with the Sachs equation~\eqref{sachs}, this implies that one can always choose a parallelly transported frame such that $L_{ij}$ takes the block diagonal form \cite{OrtPraPra09,OrtPraPra10,MalPra11,OrtPraPra13,OrtPraPra13rev}
\beqn
L_{ij}=\left(\begin {array}{cccc} \fbox{${\cal L}_{(1)}$} & & &  \\
& \ddots & & \\ 
& & \fbox{${\cal L}_{(p)}$} & \label{L_general} \\
& & & \fbox{$\begin {array}{ccc} & & \\ \ \ & \tilde{\cal L} \ \ & \\ & & \end {array}$}
\end {array}
\right) , 
\eeqn
where the first $p$ blocks are $2\times 2$ and the last block $\tilde{\cal L}$ is an $(n-2-2p)\times(n-2-2p)$-dimensional diagonal matrix. They read
\beqn
& & {\cal L}_{(\mu)}=\left(\begin {array}{cc} s_{(2\mu)} & A_{2\mu,2\mu+1} \nonumber \\
-A_{2\mu,2\mu+1} & s_{(2\mu)} 
\end {array}
\right) \qquad (\mu=1,\ldots, p) , \\
& & s_{(2\mu)}=\frac{r}{r^2+(a^0_{(2\mu)})^2} , \qquad A_{2\mu,2\mu+1}=\frac{a^0_{(2\mu)}}{r^2+(a^0_{(2\mu)})^2} , \label{s_A} \\
& &  \tilde{\cal L}=\frac{1}{r}\mbox{diag}(\underbrace{1,\ldots,1}_{(m-2p)},\underbrace{0,\ldots,0}_{(n-2-m)}) , 
\label{diagonal}
\eeqn
where $0\le 2p\le m\le n-2$, $r$ is defined in~\eqref{affine}, and the functions $a^0_{(2\mu)}$ are independent of $r$. The integer $m$ denotes the rank of $L_{ij}$ (which must be non-zero since $\bk$ is expanding), so that $L_{ij}$ is non-degenerate when $m=n-2$. 

Using the above expression for $L_{ij}$, one obtains for the optical scalars~\eqref{scalars}
\beqn
 & & (n-2)\theta=2\sum_{\mu=1}^p\frac{r}{r^2+(a^0_{(2\mu)})^2}+\frac{m-2p}{r} , \label{exp} \\
 & & \omega^2=2\sum_{\mu=1}^p\left(\frac{a^0_{(2\mu)}}{r^2+(a^0_{(2\mu)})^2}\right)^2 , \label{twist} \\
 & & \sigma^2=2\sum_{\mu=1}^p\left(\frac{r}{r^2+(a^0_{(2\mu)})^2}-\theta\right)^2+(m-2p)\left(\frac{1}{r}-\theta\right)^2+(n-2-m)\theta^2 . \label{shear} 
\eeqn

Clearly $\bk$ is twistfree ($\omega=0\Leftrightarrow A_{ij}=0$) iff $p=0$, so that non-zero twist requires $p\ge1$ and thus $m\ge2$, $n\ge4$ (cf. also \cite{ChoPoPSez10}). It is shearfree ($\sigma=0\Leftrightarrow\sigma_{ij}=0$)  when $m=n-2$ and additionally either: (i) $p=0$ (i.e., for Robinson-Trautman spacetimes \cite{PodOrt06}) or (ii) $(a^0_{(2)})^2=(a^0_{(4)})^2=\ldots=(a^0_{(2p)})^2$ and $2p=m=n-2$, which is clearly possible only when $n$ is even (cf.~\cite{OrtPraPra09,OrtPraPra10} for more details).

\subsection{Maxwell equation}

\label{subsec_Maxw}

Thanks to~\eqref{A_KS}, the non-zero components of $\bF$ read
\be
    F_{01}=D\alpha  , \qquad F_{ij}= -2\alpha A_{ij} , \qquad F_{1i}=-2\alpha L_{[1i]} - \delta_{i}\alpha  .
	\label{F_compts}		
\ee
We observe that $\bF$ is (non-zero and) {\em null} (cf., e.g., \cite{Sokolowskietal93,Coleyetal04vsi,OrtPra16}) iff $D\alpha=0=A_{ij}$. 

The Maxwell eq.~\eqref{Maxw} can thus be written as (cf., e.g., \cite{Durkeeetal10,Ortaggio14}) 
\beqn
  & & D^2\alpha + (n-2)\theta D\alpha +2\alpha \omega^2=0	,		\label{Max1} \\
  & &  \left[D\delta_i+(3A_{ji}-\sigma_{ij})\delta_j +(n-3)\theta\delta_{i}  +(L_{1i}-2L_{i1})D\right]\alpha  \nonumber \\
	& & \qquad\qquad {}+2 \alpha\Big[\delta_{j}A_{ji} +(n-4)\theta L_{[1i]}-2\sigma_{ij}L_{[1j]}-2A_{ji}L_{j1}+ \overset{k}{M}_{jj}A_{ki} +\overset{k}{M}_{ij}A_{jk}\Big]=0  ,
 		\label{Max3} \\
  & &   (\Delta D + \delta_j \delta_j+4 L_{[1j]}\delta_{j}+ N_{jj}D + \overset{k}{M}_{jj}\delta_{k}) \alpha  \nonumber \\
  & & \qquad\qquad  {}+2\alpha(2L_{[1j]}L_{[1j]}+\delta_{j}L_{[1j]}+ \overset{k}{M}_{jj}L_{[1k]} +N_{kj}A_{jk}) =0 ,
				\label{Max2}  					
\eeqn
where in~\eqref{Max3} we have used~\eqref{11e} to get rid of a term proportional to $DL_{[1i]}$. Using~\eqref{Li0_KS}--\eqref{Ni1_KS} and \eqref{covder_KS}, it is easy to see that the components~\eqref{Max1}--\eqref{Max2} of the Maxwell equation do not contain $H$ (although some of the individual terms appearing above do) and thus reduce to equations in the background space $\bet$, as expected from the comments in point~\ref{k_Maxw}. in section~\ref{subsec_backgr}.

For later use, let us note that~\eqref{Max1} can be alternatively rewritten as $D^2\ln\alpha+(D\ln\alpha)[D\ln\alpha+(n-2)\theta]+2\omega^2=0$, from which it follows
\be
 (D\ln\alpha)^2[(D\ln\alpha)^2+2\omega^2]^2=\left\{\frac{1}{2}D\left[(D\ln\alpha)^2\right]+(n-2)\theta(D\ln\alpha)^2\right\}^2 . 
 \label{Max1_alt}
\ee

\subsubsection{Twisting case ($p\ge1$)}

Eq.~\eqref{Max1} can be used to fix the $r$-dependence of $\alpha$. After some manipulations one arrives at
\be
    \alpha = \frac{\beta}{r^{m-2p-1}}\prod _{\mu=1}^{p} \frac{1}{r^2 + (a^0_{2\mu})^2}
		 \label{rescale_alpha} ,
\ee
where the auxiliary function $\beta$ is defined for even and odd $m$ as, respectively,
\beqn
   & & \beta= \alpha_{0}+  \beta_{0}\sum_{\mu=0}^p \frac{{\cal A}^0_{\mu}}{m-1-2\mu}r^{m-1-2\mu} \qquad (m\ge 2 \mbox{ even}) , \label{beta_even} \\
	 & & \beta= \alpha_{0}+  \beta_{0}\Bigg({\cal A}^0_{\frac{m-1}{2}}\ln r+ \sum_{\substack{\mu=0 \\
    (2\mu\neq m-1)}}^p \frac{{\cal A}^0_{\mu}}{m-1-2\mu}r^{m-1-2\mu}\Bigg) \qquad (m\ge 3 \mbox{ odd}) ,
\eeqn
with
\be
 {\cal A}^0_{0}=1 , \qquad\quad {\cal A}^0_{\mu}=\sum_{\nu_1<\nu_2<\ldots<\nu_\mu} (a^{0}_{(2\nu_1)})^2(a^{0}_{(2\nu_2)})^2\ldots(a^{0}_{(2\nu_{\mu})})^2	\quad (\mu=1,\ldots, p) ,
 \label{A_alpha}
\ee
and $\alpha_{0}$ and $\beta_{0}$ are two $r$-independent integration functions.

\subsubsection{Twistfree case ($p=0$)}

When there is no twist ($p=0$) a solution to~\eqref{Max1} can be written simply as
\beqn
	 & & \alpha = \alpha_{0}r^{1-m}+  \frac{\beta_{0}}{m-1} \qquad (m\neq1) ,  \label{alpha_nontwist} \\
	 & & \alpha = \alpha_{0}+\beta_{0}\ln r  \qquad (m=1) . \label{alpha_nontwist_m=1}
\eeqn

\subsection{Einstein equation}

Using~\eqref{F_compts}, one finds $F_{cd}F^{cd}=-2(D\alpha)^2+4\alpha^2\omega^2$ and thus the non-zero components of the energy-momentum tensor~\eqref{T} take the form
\beqn
  & & T_{01}=-\frac{1}{2}(D\alpha)^2-\alpha^2\omega^2 , \qquad T_{ij}=4\alpha^2A_{ik}A_{jk}-\frac{1}{2}\big[-(D\alpha)^2+2\alpha^2\omega^2\big]\delta_{ij} , \label{Tij} \\
	& & T_{1i}=-(2\alpha L_{[1i]}+\delta_i\alpha)D\alpha+2\alpha A_{ij}(2\alpha L_{[1j]}+\delta_j\alpha)  , \label{T1i} \\
	& & T_{11}=(2\alpha L_{[1i]}+\delta_i\alpha)(2\alpha L_{[1i]}+\delta_i\alpha) . \label{T11}
\eeqn

Using~\eqref{Rij} with~\eqref{OC} and \eqref{Tij}, the (spatial) trace and the tracefree part of the $(ij)$ component of the Einstein equation~\eqref{Einst} read, respectively,
\beqn 
		& & 2DH + 2H(n-2)\theta  - 2H\frac{L_{mn}L_{mn}}{(n-2)\theta} = -\frac{\kappa}{(n-2)\theta} \left[ 2\alpha^2 \omega^2 + (D\alpha)^2\right] , \label{Eii} \\ 
    & & S_{ij}\frac{(D\alpha)^2+2\alpha^2 \omega^2}{(n-2)\theta}= 4 \alpha^2 A_{ik}A_{jk} + \frac{\delta_{ij}}{n-2}\left[(D\alpha)^2 -2\alpha^2 \omega^2\right] , \label{Eij_tracefree} 
\eeqn
while the $(01)$ component is satisfied identically as a consequence of the Bianchi identity, and can thus from now on be omitted.

Owing to the block structure~\eqref{L_general}--\eqref{diagonal} of the matrix $L_{ij}$, eq.~\eqref{Eij_tracefree} contains only diagonal terms, which give rise to different conditions depending on what component one is looking at. If $p\ge1$, from the block ${\cal L}_{(1)}$ one obtains
\be
    \frac{r}{r^2 + (a^{0}_{(2)})^2}\frac{(D\alpha)^2+2\alpha^2 \omega^2}{(n-2)\theta}=4 \alpha^2 \left(\frac{a^{0}_{(2)}}{r^2 + (a^{0}_{(2)})^2}\right)^2 + \frac{1}{n-2}\left[(D\alpha)^2 -2\alpha^2 \omega^2\right] \qquad (p\ge1) ,
\label{Eij_block}		
\ee
with \eqref{exp} and \eqref{twist}, and similarly for the remaining blocks ${\cal L}_{(\mu)}$, up to ${\cal L}_{(p)}$.

Non-zero and zero entries of $\tilde{\cal L}$ are present, respectively, when $m>2p$ and $m<n-2$, and give rise to
\beqn
    & & (D\alpha)^2 -2\alpha^2 \omega^2=\frac{1}{r\theta}\left[(D\alpha)^2+2\alpha^2 \omega^2\right] \qquad\quad (m>2p) , \label{Eij_1} \\
		& & (D\alpha)^2 -2\alpha^2 \omega^2=0 \qquad\quad (m<n-2) . \label{Eij_0}
\eeqn

For the time being, we do not need to display the remaining components (i.e., the $(1i)$ and $(11)$ ones) of the Einstein equation.

\section{Twisting solutions ($n\ge4$, $p\ge1$)}

\label{sec_twisting}

As mentioned in section~\ref{subsec_opt_matr}, twist is non-zero iff $p\ge1$, which will thus be assumed throughout the present section. 
Eqs.~\eqref{Eij_1} and \eqref{Eij_0} are clearly incompatible when $\alpha^2\omega^2\neq0$, therefore we have only to consider here the following possible cases (cf.~\eqref{diagonal}): (i) $m=n-2>2p$, i.e., $\tilde{\cal L}=r^{-1}\mbox{diag}(1,\ldots,1)$; (ii) $m=n-2=2p$, i.e., the block $\tilde{\cal L}$ is absent in~\eqref{L_general} ($n$ even); (iii) $m=2p<n-2$, i.e., $\tilde{\cal L}=\mbox{diag}(0,\ldots,0)$  ($m$ even). These are analyzed in what follows. Note that necessarily $n>4$ in cases (i) and (iii), and recall that $L_{ij}$ is non-degenerate iff $m=n-2$, i.e., in cases (i) and (ii). The fact that the standard case $n=4$~\cite{DebKerSch69} is possible only in branch~(ii) is a consequence of the Goldberg-Sachs theorem \cite{GolSac62,KunTho62,RobSch63,Stephanibook}.

\subsection{Non-degenerate $L_{ij}$ with $2p<n-2$}

Combining~\eqref{Eij_block} and~\eqref{Eij_1} one obtains
\be
	(D\alpha)^2+2\alpha^2 \omega^2=-4 \alpha^2\frac{(n-2)\theta r}{r^2 + (a^{0}_{(2)})^2} ,
\ee 
which is clearly inconsistent since the LHS and the RHS have opposite signs (notice that here~\eqref{exp} gives $\theta r>0$). Therefore this case cannot occur. We observe that we arrived at this conclusion without using the Maxwell equation.

\subsection{Non-degenerate $L_{ij}$ with $2p=n-2$ ($n$ even)}

Here one has $p=1\Leftrightarrow m=2\Leftrightarrow n=4$, and since the four-dimensional case has been already elucidated \cite{DebKerSch69},\footnote{To be precise, only the case $\Lambda=0$ was studied in \cite{DebKerSch69}.} we can focus hereafter on the case $p\ge2$. Along with~\eqref{Eij_block} we thus have a similar equation with $a^{0}_{(2)}$ replaced by $a^{0}_{(4)}$. If $a^{0}_{(2)}\neq a^{0}_{(4)}$, a linear combination of those equations gives
\be
   (D\ln\alpha)^2-2\omega^2=-\frac{4(n-2)r^2}{(r^2 + (a^{0}_{(2)})^2)(r^2 + (a^{0}_{(4)})^2)} =0  \qquad (\mbox{if } a^{0}_{(2)}\neq a^{0}_{(4)}) .
	\label{Eij_24} 
\ee

Further analysis requires to consider three possible subcases separately, depending on the multiplicity of the functions $(a^{0}_{(2 \mu)})^2$.

\subsubsection{Case with at least three distinct $(a^{0}_{(2 \mu)})^2$ ($n\ge8$)}

Let us assume $p\ge3$ and that there exist at least three distinct $(a^{0}_{(2 \mu)})^2$, say (up to a relabeling of the frame vectors) $a^{0}_{(2)}$, $a^{0}_{(4)}$ and $a^{0}_{(6)}$. 
Since the LHS of~\eqref{Eij_24} is independent of the choice of the block, considering the remaining two equations obtained by performing the substitutions of indices $(24)\rightarrow(46)\rightarrow(62)$ in~\eqref{Eij_24}, one concludes that $(r^2 + (a^{0}_{(2)})^2)(r^2 + (a^{0}_{(4)})^2)=(r^2 + (a^{0}_{(4)})^2)(r^2 + (a^{0}_{(6)})^2)=(r^2 + (a^{0}_{(6)})^2)(r^2 + (a^{0}_{(2)})^2)$. This is possible only if $(a^{0}_{(2)})^2=(a^{0}_{(4)})^2=(a^{0}_{(6)})^2$, thus contradicting our assumption. We have thus proven that this case cannot occur, i.e., there can be at most two distinct $(a^{0}_{(2 \mu)})^2$.

\subsubsection{Case with two distinct $(a^{0}_{(2 \mu)})^2$ ($n\ge6$)}

Let us assume $p\ge2$ and that there exist precisely two distinct $(a^{0}_{(2 \mu)})^2$, say $a^{0}_{(2)}\neq a^{0}_{(4)}$, with respective multiplicities $p_1$ and $p_2$ (such that $2(p_1+p_2)=2p=n-2$). Eqs.~\eqref{exp} and \eqref{twist} thus give
\be
 (n-2)\theta=2r\left(\frac{p_1}{r^2+(a^0_{(2)})^2}+\frac{p_2}{r^2+(a^0_{(4)})^2}\right) , \quad \omega^2=2p_1\left(\frac{a^0_{(2)}}{r^2+(a^0_{(2)})^2}\right)^2+2p_2\left(\frac{a^0_{(4)}}{r^2+(a^0_{(4)})^2}\right)^2 .
 \label{scal_2_dist}
\ee

Substituting~\eqref{Eij_24} with~\eqref{scal_2_dist} into~\eqref{Max1_alt} leads to an inconsistency, therefore this case can also be ruled out.

\subsubsection{Shearfree case (all the $(a^{0}_{(2 \mu)})^2$ coincide)}

\label{subsubsec_shearfree}

The last non-degenerate case to consider arises when $2p=m=n-2$ and all the functions $(a^{0}_{(2 \mu)})^2$ coincide.
As mentioned in section~\ref{subsec_opt_matr}, the null vector field $\bk$ becomes then shearfree, while still expanding and twisting. In this case, no inconsistencies as the one found above arise, which means one can proceed with the full integration of the Einstein-Maxwell equations. 
According to the charging ansatz~\eqref{KS}, \eqref{A_KS}, \eqref{H_redef}, one thus needs to start from a vacuum KS solution with a shearfree twisting $\bk$, which is necessarily an mWAND (recall points~\ref{k_geod}. and \ref{k_Riem}. in section~\ref{subsec_backgr}).

For $n>4$, the first example of a Ricci-flat spacetime admitting a shearfree twisting mWAND was identified for $n=6$ in \cite{OrtPraPra13} (cf. also \cite{OrtPraPra13rev}) among the Taub-NUT vacua \cite{Berard82,BaiBat85,PagPop87} (see also \cite{Lorenz-Petzold87_nut,ChaGib96,Taylor-Robinson98,AwaCha02,ManSte04}). Subsequently, all $\Lambda=0$ vacua possessing a twisting shearfree mWAND were obtained in six dimensions in \cite{Ortaggio17}, and in $n\ge6$ (even) dimensions (including $\Lambda$) in \cite{Taghavi-Chabert21} (see also \cite{Alekseevskyetal21} for related results).\footnote{To be precise, a further condition on the asymptotic behaviour of the Weyl tensor was imposed in \cite{Ortaggio17}, which however is irrelevant from the viewpoint of the present discussion, for it is obeyed by all shearfree KS vacua\cite{OrtPraPra09}. Such a condition was not assumed in~\cite{Taghavi-Chabert21}, where the assumptions are, in fact, slightly milder than the existence of an mWAND. It is also worth noticing that the solutions of \cite{PagPop87} contain also some (non-KS) vacua for which the shearfree, twisting congruence is not an mWAND (appendix~\ref{subsubsec_NUT_generic}) or not even a WAND (appendix~\ref{subsubsec_NUTspecial}).} As it turns out \cite{Taghavi-Chabert21}, they coincide with the Taub-NUT metrics of \cite{Berard82,BaiBat85,PagPop87}. 
These are reviewed in appendix~\ref{app_NUT}, from which it follows that the only KS vacua possessing a shearfree, twisting KS vector field $\bk$ are given by the special Taub-NUT metrics~\eqref{shearfree_seed} with~\eqref{NUT_KS_cond2}. Then, using those metrics as vacuum seeds, the integration of the Einstein-Maxwell equations resulting from~\eqref{KS}, \eqref{A_KS}, \eqref{H_redef} is straightforward but lengthy, and we relegate technicalities to appendix~\ref{app_integration}.

As it turns out, the only $n>4$ charged KS solution admitting an expanding, twisting and shearfree KS vector field is given by (cf.~\eqref{shearfree}, \eqref{F=dZ_2}, \eqref{alpha_KS_NUT},  \eqref{H_KS_NUT}, \eqref{DF}, \eqref{shearfree_F}) 
\beqn
 & & \bg=\d r\otimes\bk+\bk\otimes\d r+(r^2+\ell^2)(\bh+\lambda\bk\otimes\bk)+r\frac{\mu_0-\kappa f(r)}{(r^2+\ell^2)^{\frac{n-2}{2}}}\bk\otimes\bk , \qquad \bk=\d u-2\bZ , \label{shearfree_g} \\
	& & \bF=\alpha'\d r \wedge\bk-2\alpha(r)\cF  , \qquad \cF\equiv\d\bZ	,	\label{shearfree_F2}
\eeqn
where $\bh=h_{\alpha\beta}(x)\d x^\alpha\d x^\beta$ is a K\"ahler-Einstein metric of constant holomorphic sectional curvature \cite{Bochner47,Yanobook_complex,KobNom2}, cf.~\eqref{NUT_KS_cond2} (parametrized by the coordinates $x^\alpha$ with $\alpha=1,\ldots,n-2$, denoted collectively as $x$), $\bZ=Z_\alpha(x)\d x^\alpha$ is a 1-form which lives in the base space, with K\"ahler 2-form $\cF$, and $\alpha'\equiv\d\alpha/\d r$ (while $\lambda$ was defined in~\eqref{lambda}). The functions $\alpha$ and $f$ are given by
\beqn
  & & \alpha =\frac{r}{(r^2+\ell^2)^{\frac{n-2}{2}}} \Bigg[\alpha_{0}+  \beta_{0}\sum_{\mu=0}^{\frac{n-2}{2}} \binom{\frac{n-2}{2}}{\mu} \frac{\ell^{2\mu}}{n-3-2\mu}r^{n-3-2\mu}\Bigg] , \label{alpha_KS_NUT2}	\\
  & & f'=-r^{-2}(r^2+\ell^2)^{\frac{n-4}{2}}\left[2\alpha^2\ell^2 +\textstyle{\frac{1}{n-2}}(r^2+\ell^2)^2(\alpha')^2\right] , \label{DF2}
\eeqn
and $\mu_0$, $\alpha_0$, $\beta_0$ are integration constants (see also~\eqref{Dalpha}, \eqref{DF3}). The above spacetime is of Weyl type~D. From~\eqref{R0i0j_KS}--\eqref{R1i1j_KS} it follows that for $r\to\infty$ all the components of the Weyl tensor fall off as $r^{1-n}$ or faster, which implies that these metrics are locally asymptotically (A)dS \cite{AshDas00} (or locally asymptotically flat if $\lambda=0$).

Starting form a stationary Taub-NUT ansatz, for $n>4$ such kind of solutions (but generically not in the KS class) were constructed in \cite{ManSte06} in the special case $\beta_0=0$, and in full generality in \cite{Awad06} (see also \cite{DehKoh06}). In the limit $n=4$ one recovers solutions obtained in \cite{Brill64,Carter68cmp,Ruban72}. 
While we refer to \cite{ManSte06,Awad06,DehKoh06,FloQue19} for a thorough discussion of the above solutions as well as their Euclidean counterparts, let us just mention that, for $r\to\infty$, the asymptotic behaviour of $\bg$ and $\bF$ is determined by 
\beqn
  & & 2H=-\kappa\frac{2\beta_0^2\ell^2}{(n-3)^2(n-5)}\frac{1}{r^2}+\ldots , \qquad \alpha=\frac{\beta_0}{n-3}+\frac{n-2}{(n-3)(n-5)}\frac{\beta_0\ell^2}{r^2}+\ldots \qquad (n>4) , \label{asympt_HD} \\
	& & 2H=-\frac{\mu_0}{r}+\ldots , \qquad \alpha=\beta_0+\frac{\alpha_0}{r}+\ldots \qquad (n=4) . 
\eeqn 
Elementary dimensional analysis reveals that $\mu_0$ and $\alpha_0^2$ enter $H$ at the orders $r^{3-n}$ and $r^{2(3-n)}$, respectively (while $\alpha_0$ enters $\alpha$ at the ``Coulombian'' order $r^{3-n}$, cf.~\eqref{alpha_KS_NUT2}), and thus both give rise to subleading terms when $n>4$ (and $\beta_0\neq0$). From~\eqref{shearfree_F2} and \eqref{asympt_HD} one finds that the quadratic electromagnetic invariant behaves as
\be
 F_{ab}F^{ab}=\frac{4(n-2)\beta_0^2\ell^2}{(n-3)^2}\frac{1}{r^4}+\ldots \qquad (n>4) , 
\ee
while for $n=4$ both $\beta_0^2$ and $\alpha_0^2$ contribute at order $r^{-4}$. The constants $\mu_0$, $\alpha_0$ and $\beta_0$ can thus be related to mass, electric charge and asymptotic magnetic field strength (cf. \cite{FloQue19} for related comments), respectively.

Let us further note that, in the weak-field limit, the spacetime metric is given by~\eqref{shearfree_g} with $\kappa=0$ (which corresponds to the vacuum geometry~\eqref{shearfree_seed}), and \eqref{A_KS} with \eqref{alpha_KS_NUT2} thus describes a test electromagnetic field solving the Maxwell equation in that background. For $\lambda=0=\beta_0$, the same test field solution can be obtained by using the Killing vector field $\pa_u$ as a vector potential \cite{Papapetrou66,Wald74}. In this case $H$ and $\alpha$ have the same functional form, which provides an example of the KS double copy \cite{MonOCoWhi14}, whereas this does not hold when $\beta_0\neq0$.

\paragraph{Solution for $n=6$} For definiteness, let us present the explicit form of the above solution in the case $n=6$. As discussed in appendix~\ref{subsubsec_NUT_generic}, if $n=6$ and if we assume $\lambda>0$, then the base space must be ${\mathbb C}P^2$ with the Fubini-Study metric \cite{Bochner47,Yanobook_complex,KobNom2}, namely (cf. also the real coordinates used, e.g., in \cite{GibPop78})
\be
 \bh=P^{-2}\left\{\d\rho^2+\frac{\rho^2}{4}\Big[(\d\psi+\cos\theta\d\phi)^2+P(\d\theta^2+\sin^2\theta\d\phi^2)\Big]\right\} , \qquad P=1+\frac{a^2}{6}\rho^2 ,
 \label{CP2}
\ee
where $a$ is a real constant related to the four-dimensional Ricci scalar by $\tilde R=4a^2$. Indeed metric~\eqref{CP2} satisfies~\eqref{NUT_KS_cond2} with $\lambda>0$ iff
\be
 a^2=6\lambda\ell^2 ,
  \label{CP2_fine}
\ee
which corresponds to the second of~\eqref{NUT_KS_cond}.

The 1-form $\bZ$ can be taken to be
\be
 \bZ=\ell P^{-1}\frac{\rho^2}{4}(\d\psi+\cos\theta\d\phi) ,
\ee
and the full six-dimensional KS solution is then given by~\eqref{A_KS}, \eqref{shearfree_g} with (from~\eqref{alpha_KS_NUT2}, \eqref{DF2})
\beqn
 & & \alpha=\frac{r^4}{(r^2+\ell^2)^2}\left[\frac{\alpha_0}{r^3}+\beta_0\left(\frac{1}{3}+\frac{2\ell^2}{r^2}-\frac{\ell^4}{r^4}\right)\right] , \\
 & & f=\frac{9\alpha_0^2(3 r^2+\ell^2)+96\alpha_0\beta_0\ell^2 r^3+8\beta_0^2\ell^2(-r^6+15\ell^2r^4+9\ell^4 r^2+9\ell^6)}{36r(r^2+\ell^2)^2} . 
\eeqn
(A further additive constant in $f$ has been omitted since it simply amounts to a redefinition of $\mu_0$.) In the limit $\lambda=0$ the base space becomes flat (thanks to~\eqref{CP2_fine}) and the above spacetime becomes a KS metric with a Minkowski background. The negative curvature version of the above solution (i.e., with base space $D^2$ and $\lambda<0$) is obtained by replacing $a\mapsto ia$. A similar example with ${\mathbb C}P^2$ base and $\lambda<0$ (thus not a KS metric) was presented in \cite{Awad06} (see also \cite{ManSte06,DehKoh06}).

\subsection{Degenerate $L_{ij}$ ($2p=m$)}

First, by plugging~\eqref{Eij_0} into~\eqref{Eij_block} and into the corresponding equations for $\mu=2,\ldots,p$ (i.e., with $a^{0}_{(2)}$ replaced by $a^{0}_{(2\mu)}$) one concludes that $a^0_{(2)}=a^0_{(4)}=\ldots=a^0_{2p}$. Then~\eqref{Eij_0} with~\eqref{twist} gives 
\be
 (D\ln\alpha)^2=2\omega^2=4p\left(\frac{a^0_{(2)}}{r^2+(a^0_{(2)})^2}\right)^2 ,
\ee
which is incompatible with~\eqref{Max1_alt} (to see this one should also note that, thanks to~\eqref{exp}, here $(n-2)\theta=2pr/(r^2+(a^0_{(2)})^2)$). Therefore this case cannot occur.

\section{Twistfree solutions ($p=0$)}

\label{sec_twistfree}

We now move to solutions for which $\bk$ is twistfree. In this case eq.~\eqref{Eij_block} is absent, while \eqref{Eij_1} and \eqref{Eij_0} (with $\omega=0$) imply that either $m=n-2$ (i.e., 
 $L_{ij}=S_{ij}=\theta\delta_{ij}$ is non-degenerate, with $\theta=1/r$) or $D\alpha=0$. In the former case $\bk$ is shearfree and the spacetime thus belongs to the Robinson-Trautman class, while in the latter case we can assume, without loss of generality, that $\bk$ is shearing (or else we would be again in the Robinson-Trautman branch). Let us study these two possibilities separately.

\subsection{Robinson-Trautman solutions ($m=n-2$)}

\label{subsec_RT}

Since $m=n-2$, eqs.~\eqref{alpha_nontwist}, \eqref{alpha_nontwist_m=1} give
\beqn
	 & & \alpha = \alpha_{0}r^{3-n}+  \frac{\beta_{0}}{n-3} \qquad (n\ge4) , \label{alpha_RT_n>3} \\
	 & & \alpha = \alpha_{0}+\beta_{0}\ln r  \qquad (n=3) . \label{alpha_RT_n=3}
\eeqn

The case $n=3$ has been fully explored in \cite{PodPap22} (cf. also \cite{GarciaD17} and refs. therein), while for $n=4$ we refer again the reader to \cite{DebKerSch69}. For $n\ge5$, the complete family of Robinson-Trautman electrovac solutions has been obtained in \cite{OrtPodZof08} under the assumption that $\bF$ is aligned with the Robinson-Trautman null vector field. In the case considered here, the Robinson-Trautman vector field coincides with the KS one, i.e. $\bk$, which is indeed aligned with $\bF$ by construction (eq.~\eqref{F_compts}). It follows that charged KS solutions of the Robinson-Trautman class must be a subset of the solutions of \cite{OrtPodZof08}. Among those, the only ones that are also KS reduce to the electrically charged Schwarzschild-Tangherlini metric \cite{Tangherlini63} and its extensions with hyperbolic or planar symmetry \cite{GibWil87} (this was noted in \cite{OrtPraPra09} in the vacuum case).\footnote{
This follows from the fact that for Robinson-Trautman charged metrics which are also KS, the frame spatial components $C_{ijkl}$ of the Weyl tensor fall as $r^{1-n}$ or faster as $r\to\infty$ (as follows from~\eqref{Rijkl_KS} with~\eqref{Eii}), thus forcing the base space to be of constant curvature \cite{PodOrt06,OrtPraPra13rev,OrtPodZof15}.}
Without the need of any further calculations we thus arrive at 
\beqn
  & & \bg=2\d u\d r+r^2\bh-\left(K-\lambda r^2-\frac{\mu_0}{r^{n-3}} + \kappa\frac{(n-3)\alpha_0^2}{n-2}\frac{1}{r^{2(n-3)}} \right)\d u^2 , \label{RT_g} \\
	& & \bF= -\frac{(n-3)\alpha_0}{r^{n-2}}\d r \wedge \d u  ,
 \label{RT_F}
\eeqn
where $\lambda$, $\mu_0$ and $\alpha_0$ are constants proportional, respectively, to $\Lambda$ (cf.~\eqref{lambda}), mass and electric charge, and $\bh=h_{\alpha\beta}(x)\d x^\alpha\d x^\beta$ is an $(n-2)$-dimensional Riemannian metric of constant curvature, whose Ricci scalar is normalized as ${\cal R}=K(D-2)(D-3)$ with $K=\pm1, 0$ ($\beta_0$ in~\eqref{alpha_RT_n>3} becomes also a constant in this case and thus a purely gauge term). Here $\bk=\d u$, $2H=-\mu_0 r^{3-n} + \kappa\frac{(n-3)\alpha_0^2}{n-2}r^{2(3-n)}$, and the constant curvature background metric $\bet$ is obtained by setting $\mu_0=0=\alpha_0$. For an appropriate parameter range, these solutions describe electrically charged black holes \cite{KodIsh04}. These spacetimes are of Weyl type~D.

Similarly as in section~\ref{subsubsec_shearfree}, the weak-field limit corresponds to setting $\kappa=0$ in~\eqref{RT_g}, in which case \eqref{RT_F} (i.e., $\bA=\alpha_0r^{3-n}\bk$) represents a test electric field living in an Einstein spacetime. For $\lambda=0$, it can be produced with the method of \cite{Papapetrou66,Wald74} owing to the presence of a Killing vector field $\pa_u$. It was discussed in the context of the KS double copy in \cite{MonOCoWhi14} (see also \cite{BahLunWhi17,CarPenTor18} for the case $\lambda\neq0$, and \cite{Aliev07_2} for related comments).

Let us further note that the Robinson-Trautman solutions of \cite{OrtPraPra09} contain an additional magnetic branch (in even dimensions), which does not belong to the charged KS class because it has $F_{ij}\neq0$ and yet $A_{ij}=0$, which is clearly incompatible with~\eqref{F_compts} (in other words, in that case $\bA$ must contain also a spatial component and cannot thus be simply proportional to $\bk$, as we have assumed). Indeed $F_{ij}=0$ in~\eqref{RT_F}.

\subsection{Shearing solutions with $D\alpha=0$ ($n\ge5$)}

\label{subsec_shearing_null}

For $n=3$ all null geodesic congruences are shearfree \cite{ChoPoPSez10}, while for $n=4$ the same is true for KS congruences, as a consequence of the Goldberg-Sachs theorem \cite{GolSac62,KunTho62,RobSch63,Stephanibook} (cf. also \cite{DebKerSch69}). Therefore here we can assume $n\ge5$.

For this class of solutions one simply has
\be
 \alpha = \frac{\beta_{0}}{m-1} ,
\ee
which is compatible with~\eqref{alpha_nontwist}, \eqref{alpha_nontwist_m=1}. Since this is the only case corresponding to a {\em null} Maxwell field (cf. section~\ref{subsec_Maxw}), this branch will be studied in detail elsewhere. Here we only point out that it is not empty by presenting an explicit five-dimensional example in what follows.

\paragraph{An example for $n=5$ ($\lambda=0$)}

A simple solution with $\lambda=0$ in five-dimension (and $m=2$) can be constructed by taking a direct product of a Vaidya-like spacetime obtained in \cite{RobTra62} (see also related comments in~\cite{Senovilla15}) with a flat extra dimension, which gives rise to
\beqn
  & & \bg=2\d u\d r+r^2(\d x^2+\d y^2)+\d z^2+\frac{\mu_0+h(u)}{r}\d u^2 , \qquad \dot h=\kappa(a_1^2+a_2^2), \label{string_g} \\
	& & \bF= \big[a_1(u)\d x+a_2(u)\d y\big]\wedge \d u  ,  \label{string_F}
\eeqn
where $\mu_0$ is a constant, $a_1$, $a_2$ are arbitrary functions of the advanced time $u$, and $\dot h\equiv\d h/\d u$. The radiative (null) field \eqref{string_F} (or \eqref{A_KS} with $\alpha=a_1(u)x+a_2(u)y$) is responsible for the time dependence of line-element and corresponds to a pure radiation energy-momentum tensor given by 
\be
	\bT=\frac{a_1^2+a_2^2}{r^2}\d u^2 .
\ee	
This can be interpreted as an energy flux along $k^a\pa_a=\pa_r$, which produces (loosely speaking) a mass gain due to incoming radiation as $u$ evolves.\footnote{Alternatively, after the replacement $u\mapsto-u$, one can interpret the solution as describing emission of radiation (and mass loss) as the retarded time $u$ evolves (cf., e.g., \cite{Senovilla15}).} We have checked that the above spacetime does not admit any mWAND distinct from $\bk$ and is therefore of Weyl type~II. However, in regions where $r(h+\mu_0)<0$ there exists a unique Weyl aligned null direction (WAND) of multiplicity 1 \cite{Coleyetal04,OrtPraPra13rev} defined by $\pa_u-\frac{1}{2}\left(\zeta^2+\frac{\mu_0+h}{r}\right)\pa_r+\zeta\pa_z$ with $\zeta^2=-\frac{r\dot h}{3(h+\mu_0)}$. Metric~\eqref{string_g} is flat iff $a_1=0=a_2$ and $\mu_0=0$.

\section*{Acknowledgments}

This work has been supported  by the Institute of Mathematics, Czech Academy of Sciences (RVO 67985840).

\renewcommand{\thesection}{\Alph{section}}
\setcounter{section}{0}

\renewcommand{\theequation}{{\thesection}\arabic{equation}}

\section{The Kerr-Newman-(A)dS solution in KS form ($n=4$)}
\setcounter{equation}{0}

\label{app_KNAdS}

\subsection{Spherical solution}

Similarly as described in section~\ref{intro} in the Kerr-Newman case, the four-dimensional Kerr-Newman-(A)dS solution in KS coordinates can be obtained with the same method by ``charging'' the Kerr-(A)dS metric. Starting from the vacuum line-element in KS coordinates given in  \cite{Gibbonsetal05}, one easily arrives at the Einstein-Maxwell solution 
\beqn
\bg=\bet+\frac{2mr-e^2}{\rho^2}\bk\otimes\bk , 
\label{KNAdS}
\eeqn
where\footnote{Cf. \cite{HawHunTay99} for the special case $\Xi=0$ (for $\lambda<0$).}
\beqn
 & & \bet=-\frac{\Delta_\theta}{\Xi}(1-\lambda r^2)\d t^2+\frac{\rho^2}{(1-\lambda r^2)(r^2+a^2)}\d r^2+\frac{\rho^2}{\Delta_\theta}\d\theta^2+\frac{r^2+a^2}{\Xi}\sin^2\theta\d\phi^2 , \\
 & & \Delta_\theta=1 + \lambda a^2\cos^2\theta , \qquad \rho^2=r^2+a^2\cos^2\theta , \qquad \Xi=1 + \lambda a^2 , \qquad \lambda=\frac{\Lambda}{3} , \label{KNAdS_functs}
\eeqn
is a suitable form of the (A)dS metric, and 
\be
 \bk=\frac{\Delta_\theta}{\Xi}\d t+\frac{\rho^2}{(1-\lambda r^2)(r^2+a^2)}\d r-\frac{a\sin^2\theta}{\Xi}\d\phi ,
  \label{k_KNAdS}
\ee
represents an affinely parametrized geodesic null vector field.
The vector potential corresponding to metric~\eqref{KNAdS} is given by\footnote{We observe that, in the test-field limit, the KS form of the solution provides a natural rephrasing of the Killing 1-form ``background subtraction'' of \cite{Aliev07_2}.} 
\be
 \bA=-\frac{er}{\rho^2}\bk. 
 \label{KNAdS_A}
\ee 
For $\lambda=0$ the above solution reduces to~\eqref{KS}--\eqref{A_KN} upon redefining the coordinates as $\d t\mapsto\d u-\d r$, $\d\phi\mapsto-\d\phi-a\d r/(r^2+a^2)$.

The Kerr-Newman-(A)dS solution~\eqref{KNAdS}--\eqref{KNAdS_A} is usually presented in Boyer-Lindquist-type coordinates. These are defined by (cf. \cite{Gibbonsetal05} in the vacuum case)
\beqn
 & & \d t=\frac{\d\tau}{\Xi}+\frac{2mr-e^2}{(1-\lambda r^2)\Delta_r}\d r , \qquad \d\phi=\d\varphi-a\lambda\frac{\d\tau}{\Xi}+a\frac{2mr-e^2}{(r^2+a^2)\Delta_r}\d r , \label{BL} \\
 & & \Delta_r=\left(a^2+r^2\right) \left(1-\lambda  r^2\right)-2 m r+e^2 , 
\eeqn 
resulting in
\be
 \bg=-\frac{\Delta_r}{\Xi^2\rho^2}\left(\d\tau-a\sin^2\theta\d\varphi\right)^2+\frac{\Delta_\theta\sin^2\theta}{\Xi^2\rho^2}\left[a\d\tau-(r^2+a^2)\d\varphi\right]^2+\frac{\rho^2}{\Delta_r}\d r^2+\frac{\rho^2}{\Delta_\theta}\d\theta^2 , \label{KNAdSBL} 
\ee
with~\eqref{KNAdS_functs}, \eqref{KNAdS_A}, and \eqref{k_KNAdS} taking the form
\be
 \bk=\frac{\d\tau}{\Xi}+\frac{\rho^2}{\Delta_r}\d r-\frac{a\sin^2\theta}{\Xi}\d\varphi .
 \label{kBL}
\ee
After observing that the term proportional to $\d r$ in~\eqref{kBL} gives rise to a removable gauge term in~\eqref{KNAdS_A}, this is the form of the solution given in \cite{Carter68cmp,Carter73} (cf. also \cite{GriPodbook}).

\subsection{Hyperbolic solution}

In addition to the Kerr-Newman-(A)dS solutions considered above, the family of metrics obtained by Carter \cite{Carter68pla,Carter68cmp} and Pleba\'nski \cite{Plebanski75_2} (see also \cite{Carter73,PleDem76}) also contains certain ``topological'' extensions, as described in \cite{KleMorVan98,CalKle99,Klemm98}. The hyperbolic counterpart of \eqref{KNAdS}--\eqref{KNAdS_A} can be obtained straightforwardly by the following analytic continuation of coordinates and parameters \cite{KleMorVan98,CalKle99}
\begin{align}
&t \mapsto  it, \qquad r\mapsto  ir, \qquad \theta \mapsto  i\theta, \nonumber\\
& m\mapsto  -im , \qquad a\mapsto  i a .
\end{align}
The resulting solution is given again by~\eqref{KNAdS} and \eqref{KNAdS_A}, but now with (after redefining $\rho\mapsto i\rho$, $\bk\mapsto-i\bk$)
\begin{align}
& \bet=-\frac{\Delta_\theta}{\Xi}(-1-\lambda r^2)\d t^2+\frac{\rho^2}{(-1-\lambda r^2)(r^2+a^2)}\d r^2+\frac{\rho^2}{\Delta_\theta}\d\theta^2+\frac{r^2+a^2}{\Xi}\sinh^2\theta\d\phi^2 , \label{KNAds_eta_hyp}\\ 
 & \Delta_\theta=1 - \lambda a^2\cosh^2\theta , \qquad \rho^2=r^2+a^2\cosh^2\theta , \qquad \Xi=1 - \lambda a^2 , 
	\label{KNAd_hyp_metfns} \\
& \bk=\frac{\Delta_\theta}{\Xi}\d t+\frac{\rho^2}{(1+\lambda r^2)(r^2+a^2)}\d r+\frac{a\sinh^2\theta}{\Xi}\d\phi . \label{KNAds_k_hype}
\end{align}

The solutions of \cite{KleMorVan98,CalKle99} were not given in the above KS form, but in Boyer-Lindquist-type coordinates -- these can be obtained by an obvious modification of~\eqref{BL}.

\subsection{Planar solution}

The KS form of the planar solution of \cite{KleMorVan98,CalKle99} is given by~\eqref{KNAdS} and \eqref{KNAdS_A} with
\begin{align}
    &\bet = \frac{N }{\rho^2}\d t^2 + \frac{\rho^2}{a^2- \lambda r^4}\d r^2 + \frac{\rho^2}{\Delta_P} \d P^2 - \frac{2aS}{\rho^2}\d t \d \phi  + (r^2-a^2P^2) \d \phi^2  , \\
    &  \bk= (1- \lambda a^2 P^2) \d t + \frac{\rho^2}{a^2- \lambda r^4} \d r + a P^2 \d \phi,
    \end{align}
where
    \begin{align}
    & \Delta_P =1-  \lambda a^2  P^4 , \qquad \rho^2= r^2 + a^2P^2, \nonumber\\
    & N =   a^2(1+\lambda r^2 )^2 \Delta_P - (a^2-\lambda r^4)(1-\lambda a^2 P^2)^2,  \\
    & S=  P^2 (a^2-\lambda r^4)(1-\lambda a^2 P^2) + r^2 (1 + \lambda r^2) \Delta_P .  \nonumber
\end{align}

As in the hyperbolic case, the solutions of \cite{KleMorVan98,CalKle99} were presented in Boyer-Lindquist-type coordinates, which can be obtained by defining
\beqn
 & & \d t = \d\tau + \frac{(2mr-e^2) r^2}{(a^2- \lambda r^4)\Delta_r} \d r , \qquad \d\phi = \d \varphi + a \lambda \d t + a \frac{2mr-e^2}{(a^2-\lambda r^4)\Delta_r}\d r , \\
 & & \Delta_r= a^2 -  \lambda r^4 - 2m r +e^2 .
\eeqn

\section{Connection and curvature of KS spacetimes ($\bk$ geodesic)}
\setcounter{equation}{0}

\label{app_Riem}

The Ricci rotation coefficients~\eqref{Ricci_rot} for the geometry~\eqref{KS} in a null frame $\{\bk,\bn,\bm_{(i)}\}$ adapted to $\bk$ (as defined in section~\ref{subsec_notat}) read
\beqn
 & & L_{i0}=L_{i0}|_{0} , \qquad L_{10}=L_{10}|_{0} , \qquad L_{ij}=L_{ij}|_{0} , \qquad \M{i}{j}{0}=\M{i}{j}{0}|_{0} , \qquad \M{i}{j}{k}=\M{i}{j}{k}|_{0} \label{Li0_KS} \\
 & & N_{i0}=N_{i0}|_{0} , \qquad L_{i1}=L_{i1}|_{0} , \qquad L_{1i}=L_{1i}|_{0}-HL_{i0}|_{0} , \qquad N_{ij}=N_{ij}|_{0}+HL_{ji}|_{0} , \label{Nij_KS} \\
 & & \M{i}{j}{1}=\M{i}{j}{1}|_{0}+H\big(\M{i}{j}{0}|_{0}+2L_{[ij]}|_{0}\big) , \qquad  L_{11}=L_{11}|_{0}-HL_{10}|_{0}-D|_{0}H , \label{L11_KS} \\
 & & N_{i1}=N_{i1}|_{0}+H\big(N_{i0}|_{0}+2L_{1i}|_{0}-HL_{i0}|_{0}-L_{i1}|_{0}\big)+\delta_i|_{0}H ,  \label{Ni1_KS}
\eeqn 
where a subscript $|_{0}$ denotes quantities pertaining to the background geometry $\bet$ (defined by $H=0$). We note that the frames of $\bg$ and $\bet$ are related by $k^a=k^a|_{0}$, $m_{(i)}^a=m_{(i)}^a|_{0}$ and $n^a=n^a|_{0}+Hk^a|_{0}$ (and thus $n_a=n_a|_{0}-Hk_a|_{0}$), so that for the derivative operators~\eqref{covder}, when acting on scalar functions, we have 
\be
 D=D|_{0} , \qquad \delta_i=\delta_i|_{0} , \qquad \Delta=\Delta|_{0}+HD|_{0} .
 \label{covder_KS}
\ee

When $\bk$ is {\em geodesic} ($\Leftrightarrow L_{i0}=0$, which is necessarily the case, e.g., in vacuum or with aligned matter \cite{OrtPraPra09,MalPra11}) and an {\em affine parametrization} is employed ($\Leftrightarrow L_{10}=0$), and $\bet$ is a spacetime of constant curvature, the curvature tensor of metric~\eqref{KS} takes the form (with~\eqref{lambda}) \cite{OrtPraPra09,MalPra11}\footnote{We emphasize that the following expressions hold in any null frame adapted to $\bk$, i.e., not necessarily a parallelly transported one.}
\beqn
	& & R_{0i0j} = 0 , \qquad R_{010i} = 0 , \qquad R_{0ijk} = 0 , \label{R0i0j_KS} \\
	& & R_{0101} = D^2 H - \lambda , \qquad R_{01ij} = - 2 A_{ij}D H + 4 H S_{k[j} A_{i]k} , \label{R01ij_KS2} \\
	& & R_{0i1j} = - L_{ij} D H + 2  H A_{ik} L_{kj}
           + \lambda\delta_{ij} , \\
	& & R_{ijkl} = 4  H \left( A_{ij} A_{kl} + A_{l[i} A_{j]k} + S_{l[i} S_{j]k} \right) + 2\lambda\delta_{i[k}\delta_{l]j} , \label{Rijkl_KS} \\
  & & R_{011i} = \big(-\delta_i D  + 2 L_{[i1]}D + L_{ji} \delta_j\big) H + 2  H \left( L_{1j} L_{ji} - L_{j1} S_{ij} \right)  , \\
  & &  R_{1ijk}=2\big(L_{[j|i}\delta_{|k]} +A_{jk}\delta_i\big) H  \nonumber \\
  & & \qquad\qquad  {}-2 H\big(\delta_iA_{kj}+2L_{1[j}L_{k]i}{}-2L_{[j|1}A_{|k]i}+2L_{[1i]}A_{kj}+2A_{l[j}\M{l}{k]}{i}\big) , \label{R1ijk_KS} \\
	& &  R_{1i1j}=\big[\delta_{(i}\delta_{j)}+\M{k}{(i}{j)}\delta_k +2(2L_{1(j|}-L_{(j|1})\delta_{|i)}+N_{(ij)}D -S_{ij}\Delta\big]H  \nonumber \\
  & & \qquad\qquad {}+2 H\big(\delta_{(i|}L_{1|j)}-\Delta S_{ij}-2L_{1(i}L_{j)1}+2L_{1i}L_{1j}-L_{k(i|}N_{k|j)}\nonumber \\
	& & \qquad\qquad {}+L_{1k}\M{k}{(i}{j)}-2S_{k(i}\M{k}{j)}{1}-2 H S_{k(i}A_{j)k}\big) . \label{R1i1j_KS} 
\eeqn

It follows that the non-vanishing  components of the Ricci tensor are \cite{OrtPraPra09,MalPra11}
\beqn
 R_{01} & = & -[D^2 H+(n-2)\theta D H+2 H\omega^2]+(n-1)\lambda , \label{R01} \\
 R_{ij} & = & 2 H L_{ik}L_{jk}-2[D H+(n-2)\theta H]S_{ij}+(n-1)\lambda\delta_{ij} , \label{Rij} \\
 R_{1i} & = & \big(-\delta_iD +2L_{[i1]}D +2L_{ij}\delta_j -L_{jj}\delta_i\big) H \nonumber \\ 
			  & &  {}+2 H\big(\delta_jA_{ij}+A_{ij}\M{j}{k}{k}-A_{kj}\M{i}{k}{j}-L_{jj}L_{1i}+3L_{ij}L_{[1j]}+L_{ji}L_{(1j)}\big) , \label{R1i} \\  
 R_{11} & = & \big[\delta_i\delta_i +(N_{ii}-2 H L_{ii})D +(4L_{1j}-2L_{j1}-\M{i}{j}{i})\delta_j -L_{ii}\Delta\big] H \nonumber \\
				& & {}+2 H\big[2\delta_iL_{[1i]}+4L_{1i}L_{[1i]}+L_{i1}L_{i1}-L_{11}L_{ii} \nonumber \\ 
				& & {}+2L_{[1j]}\M{j}{i}{i}-2A_{ij}N_{ij}-2 H\omega^2+(n-2)\lambda\big] . \label{R11} 
\eeqn
(The Ricci identity \cite{OrtPraPra07} has also been used in~\eqref{R11}.) We observe that the Ricci tensor is linear in $H$ (this is manifest for the components~\eqref{R01}--\eqref{R1i}, while in~\eqref{R11} one has to use~\eqref{Nij_KS} and~\eqref{covder_KS}).  A similar property of the mixed coordinate components $R^a_{\phantom{a}b}$ was pointed out in \cite{GurGur75,DerGur86}.

For later purposes, it is useful to observe that, when $\bk$ is shearfree, the second component in~\eqref{R01ij_KS2} reduces to
\be
 R_{01ij}=2A_{ij}(-DH+2H\theta) \qquad (\sigma=0) .
 \label{R01ij_KS}
\ee 
This implies that if $A_{ij}\neq0$, then necessarily $R_{01ij}\neq0$ for vacuum solutions (as can be seen easily using~\eqref{exp} and the form of $H$ in vacuum obtained in\cite{OrtPraPra09,MalPra11}), except in the trivial case $H=0$. This observation will be used in section~\ref{subsec_NUT_vac}.

For $n=4$, equivalent results were obtained in \cite{Edelen66_I,Dozmorov70}.

\section{Taub-NUT spacetimes in higher dimensions}
\setcounter{equation}{0}

\label{app_NUT}

The Taub-NUT spacetime is a well known vacuum solution of Einstein gravity in four dimensions \cite{Taub51,NewTamUnt63,Ruban72} (cf. also \cite{Stephanibook,GriPodbook} and refs. therein). Extensions to arbitrary even dimensions have been known for some time \cite{Berard82,BaiBat85,PagPop87,Lorenz-Petzold87_nut,ChaGib96,Taylor-Robinson98,AwaCha02,ManSte04}. More recently, those have been characterized by the presence of two twisting shearfree congruences of null geodesics and have thus attracted attention in the context of higher-dimensional formulations of the Goldberg-Sachs theorem \cite{OrtPraPra13,OrtPraPra13rev,Ortaggio17,Taghavi-Chabert21}. In this appendix we review the basic properties of such  spacetimes, first ``off-shell'' and subsequently in the Einstein (vacuum) case (relevant to section~\ref{subsubsec_shearfree}). We also present a few new observations regarding the Weyl type and the overlap with the KS class, and further point out that Taub-NUT Einstein spacetimes (and thus shearfree congruences of null geodesics) also exist in odd dimensions in the special branch of section~\ref{subsubsec_NUTspecial}. In this appendix, we will denote by $\d s^2$ those line-elements that are not necessarily KS (as opposed to the symbol $\bg$ used for KS metrics throughout the paper).

\subsection{Metric ansatz and off-shell properties}

\subsubsection{Line-element in stationary (NUT) coordinates}

We consider the $n$-dimensional stationary line-element \cite{BaiBat85,PagPop87}
\be
 \d s^2=-A^2(r)(\d t-2\bZ)^2+B^2(r)\d r^2+C^2(r)\bh ,
 \label{NUT}
\ee
where the coordinates $x^\alpha$ ($\alpha=1,\ldots,n-2$, also denoted collectively simply as $x$) parametrize a Riemannian base space of dimension $n-2$ which carries a metric $\bh=h_{\alpha\beta}(x)\d x^\alpha\d x^\beta$, $\bZ=Z_\alpha(x)\d x^\alpha$ is a 1-form which lives in the base space, and $A$, $B$ and $C$ are functions of $r$.

For later purposes, it is useful to define the (purely spatial) 2-form
\be
 \cF\equiv\d\bZ ,
 \label{F=dZ}
\ee
which we assume to be non-zero, as well as its positive definite quadratic invariant
\be
 {\cal F}^2\equiv h^{\alpha\gamma}h^{\beta\delta}{\cal F}_{\alpha\beta}{\cal F}_{\gamma\delta} .
  \label{F2}
\ee
Both the above quantities are $r$-independent.

\subsubsection{Line-element in null coordinates and twisting, shearfree null congruences}

The two 1-forms 
\be
 \bl_{\pm}\equiv\d t-2\bZ\pm BA^{-1}\d r ,
 \label{NUT_wands}
\ee
define two congruences of {\em shearfree} null geodesics \cite{OrtPraPra13,Ortaggio17,Alekseevskyetal21,Taghavi-Chabert21} with expansion and twist given by, respectively (cf. also~\eqref{Lij_NUT} below; hereafter a prime denotes differentiation w.r.t. $r$)
\be
 \theta=\pm C'(ABC)^{-1} , \qquad \omega^2=C^{-4}{\cal F}^2 .
\ee
Note that the ``reflection'' $t\mapsto-t$, $\bZ\mapsto-\bZ$ leaves the metric invariant while sending $\bl_+\mapsto-\bl_-$ and $\bl_-\mapsto-\bl_+$. This means these two null directions have identical geometric properties (up to certain signs) \cite{PraPraOrt07}.

The change of coordinates
\be
 \d t=\d u-\frac{B}{A}\d r ,
 \label{NUT_u}
\ee
puts metric~\eqref{NUT} in the form $\d s^2=-A^2(\d u-2\bZ)^2+2AB(\d u-2\bZ)\d r+C^2(r)\bh$, and gives $\bl_+=\d u-2\bZ$ , $\bl_-=\bl_+-2BA^{-1}\d r$. Without losing generality, a redefinition of $r$ enables one to set
\be
 AB=1 ,
 \label{gauge_NUT}
\ee
which we shall assume hereafter. After relabeling
\be
 2\H=A^2 ,
\ee
we thus arrive at 
\be
 \d s^2=\d r\otimes\bl_++\bl_+\otimes\d r+C^2(r)\bh -2\H(r)\bl_+\otimes\bl_+ , \qquad \bl_+=\d u-2\bZ ,
 \label{NUT2}
\ee
where the function $\H$ can now have any sign or vanish (thus describing also possible time-dependent Taub regions).

Let us introduce the following null coframe
\be
 \bo^0=-\H\bl_- , \qquad \bo^1=\bl_+ , \qquad \bo^i=C\bot^i \qquad (i=2,\ldots,n-1) , 
 \label{coframe}
\ee
where $\{\bot^i\}$ is an orthonormal coframe of the metric $\bh$, such that $\bh=\delta_{\tilde i\tilde j}\bo^{\tilde i}\bo^{\tilde j}$ and $\d s^2=\bo^0\bo^1+\bo^1\bo^0+\delta_{ij}\bo^i\bo^j$. In the frame dual to~\eqref{coframe} (for which, in particular, $\bE_0=\ell_+^a\pa_a=\pa_r$), the non-zero Ricci rotation coefficients~\eqref{Ricci_rot} read
\beqn
 & & L_{ij}=C'C^{-1}\delta_{ij}+C^{-2}{\cal F}_{\tilde i\tilde j} , \qquad \M{i}{j}{0}=-C^{-2}{\cal F}_{\tilde i\tilde j}  \qquad \M{i}{j}{k}=-C^{-1}\Gamma^{\tilde i}_{\phantom{\tilde i}\tilde j\tilde k} , \label{Lij_NUT} \\
 & & N_{ij}=\H L_{ji} , \qquad \M{i}{j}{1}=\H C^{-2}{\cal F}_{\tilde i\tilde j} , \qquad  L_{11}=-\H' , \label{Nij_NUT}
\eeqn 
where hereafter indices with a tilde denote components in the tilded frame and $\Gamma^{\tilde i}_{\phantom{\tilde i}\tilde j\tilde k}$ are the connection coefficients of the geometry $\bh$. In particular, the twist matrix of $\bl_+$ is thus given by
\be
  A_{ij}=C^{-2}{\cal F}_{\tilde i\tilde j} .
	\label{NUT_twist} 
\ee

We note that, since $\M{i}{j}{0}\neq0$, the above frame is {\em not} parallelly transported along the null congruence generated by $\bl_+$.

\subsubsection{Curvature}

\label{subsubsec_NUT_curvat}

The non-zero frame components of the Riemann tensor can be computed straightforwardly and take the form (cf. also~\cite{PagPop87,Taghavi-Chabert21})
\beqn
  & & R_{0i0j}=C^{-4}\big(-C^3C''\delta_{ij}+{\cal F}_{\tilde i\tilde k}{\cal F}_{\tilde j\tilde k}\big) , \qquad R_{0ijk}= -2C^{-3}\tilde\nabla_{[\tilde k}{\cal F}_{\tilde j]\tilde i} , \label{R0i0j} \\
	& & R_{01ij}=-\big(2\H C^{-2})'{\cal F}_{\tilde i\tilde j} , \qquad R_{0101}=\H'' ,  \label{R01ij} \\
	& & R_{0i1j}=-\big(\H C^{-2})'{\cal F}_{\tilde i\tilde j}-C^{-1}\big(\H C'\big)'\delta_{ij}-C^{-4}\H{\cal F}_{\tilde i\tilde k}{\cal F}_{\tilde j\tilde k} , \\ 
	& & R_{ijkl}=C^{-2}\tilde R_{\tilde i\tilde j\tilde k\tilde l}+4\H C ^{-4}\big[-\big(CC'\big)^2\delta_{i[k}\delta_{l]j}+{\cal F}_{\tilde i\tilde j}{\cal F}_{\tilde k\tilde l}-{\cal F}_{\tilde i[\tilde k}{\cal F}_{\tilde l]\tilde j}\big] , \label{Rijkl} \\
	& & R_{1i1j}=\H^2R_{0i0j}  , \qquad R_{1ijk}=-\H R_{0ijk} , 
\eeqn
where $\tilde\nabla$ and $\tilde R_{\tilde i\tilde j\tilde k\tilde l}$ are, respectively, the covariant derivative and the Riemann tensor associated to $\bh$. In passing, let us observe that, by the first of~\eqref{R0i0j}, $\bl_{\pm}$ are WANDs iff $(n-2){\cal F}_{\tilde i\tilde k}{\cal F}_{\tilde j\tilde k}={\cal F}^2\delta_{\tilde i\tilde j}$ (which is an identity for $n=4$).

The Ricci components then follow (cf. also~\cite{Berard82,BaiBat85,PagPop87,Taghavi-Chabert21})
\beqn
  & & R_{00}=C^{-4}\big[-(n-2)C^3C''+{\cal F}^2\big] , \qquad R_{0i}= C^{-3}\tilde\nabla_{\tilde k}{\cal F}_{\tilde k\tilde i} , \\	
	& & R_{01}=-\H''-(n-2)C^{-1}\big(\H C'\big)'-C^{-4}\H{\cal F}_{\tilde i\tilde k}{\cal F}_{\tilde i\tilde k} ,  \\
	& & R_{ij}=C^{-2}\big(\tilde R_{\tilde i\tilde j}+4\H C ^{-2}{\cal F}_{\tilde i\tilde k}{\cal F}_{\tilde j\tilde k}\big)-2\delta_{ij}C^{-2}\big[C(\H C')'+(n-3)\H(C')^2\big] , \\	
	& & R_{11}=\H^2R_{00}  , \qquad R_{1i}=-\H R_{0i} . 
\eeqn

\subsection{Vacuum solutions}

\label{subsec_NUT_vac}

For the purposes of this paper, we are interested in Taub-NUT metrics that solve the $\Lambda$-vacuum Einstein equation
\be
 R_{ab}=\frac{R}{n}g_{ab} , \label{Einst_vac} \\ 
\ee
which gives (cf.~\eqref{Einst} with $T_{ab}=0$ and \eqref{lambda}) 
\be
 R=n(n-1)\lambda .
\ee

Imposing $R_{00}=0$ gives $(n-2)C^3C''={\cal F}^2$. Since $C$ depends only on $r$, this means that ${\cal F}^2=$const. The solution of this ODE can be written (up to a linear redefinition of $r$ and suitable rescalings of $u$ and $\bZ$, such as to preserve~\eqref{gauge_NUT}) as \cite{BaiBat85,PagPop87}
\be
 C^2=r^2+\ell^2 , \qquad \ell^2\equiv \frac{{\cal F}^2}{n-2}(=\mbox{const}) .
 \label{C_canonic}
\ee

From $R_{0i}=0$ one obtains
\be
 \tilde\nabla_{\tilde k}{\cal F}_{\tilde k\tilde i}=0 ,
\ee
so that $\cF$ is also co-closed (in addition to being closed, cf.~\eqref{F=dZ}). 

In order to solve $R_{ij}=\frac{R}{n}\delta_{ij}$ one needs to consider two possibilities separately, depending on whether or not $\H C^{-2}$ is a constant.\footnote{This was noticed already in \cite{PagPop87}, but the special case $\H C^{-2}=$const (section~\ref{subsubsec_NUTspecial}) was not studied there.}

\subsubsection{Generic case $\H C^{-2}\neq$const ($n$ even)}

\label{subsubsec_NUT_generic}

In this case, requiring $R_{ij}=\frac{R}{n}\delta_{ij}$ gives rise to three separate equations. Two of those are tensorial and read
\beqn
 & & {\cal F}_{\tilde i\tilde k}{\cal F}_{\tilde j\tilde k}=\ell^2\delta_{ij} , \label{FF_Kahl} \\
 & & \tilde R_{\tilde i\tilde j}=\frac{\tilde R}{n-2}\delta_{ij} , \label{Rij_base}
\eeqn 
where $\tilde R_{\tilde i\tilde j}=\tilde R_{\tilde i\tilde k\tilde j\tilde k}$ and $\tilde R=\tilde R_{\tilde k\tilde k}$. This means that the spatial geometry must be (almost-)K\"ahler-Einstein \cite{PagPop87,Taghavi-Chabert21} and  $n$ is necessarily even. We further observe that~\eqref{FF_Kahl} ensures that the two null directions defined by~\eqref{NUT_wands}
are WANDs (cf. section~\ref{subsubsec_NUT_curvat}). They are mWANDs (i.e., the Weyl type is D) if, and only if, $\tilde\nabla_{\tilde k}{\cal F}_{\tilde i\tilde j}=0$, i.e., if the base space is K\"ahler-Einstein (cf. also \cite{Taghavi-Chabert21}).

The remaining (scalar) equation determines $\H(r)$ up to an arbitrary integration constant, and can be conveniently written as \cite{PagPop87}
\be
 \frac{r^2}{(r^2+\ell^2)^{n/2}}\left[r^{-1}(r^2+\ell^2)^{(n-2)/2}2\H\right]'+\frac{1}{n-2}\left(2\Lambda-\frac{\tilde R}{r^2+\ell^2}\right)=0 .
 \label{NUT_eq}
\ee
Note that $2\H(r^2+\ell^2)^{(n-2)/2}$ is a polynomial of degree $n$ in $r$ \cite{PagPop87,ChaGib96,Alekseevskyetal21}. For large values of $r$ this gives the asymptotic behaviour 
\be
 2\H=-\lambda r^2+\frac{-\lambda\ell^2 (2n-3)+\frac{\tilde R}{n-2}}{n-3}+\ldots ,
\ee
with an integration constant $\mu_0$ appearing at the order $r^{3-n}$.

Upon using the Bianchi identity, one can then verify that all components of the Einstein equation are now satisfied. The resulting vacuum metric is thus given by~\eqref{NUT2} with \eqref{F=dZ} and \eqref{C_canonic}--\eqref{NUT_eq}. In the limit $\ell\to0$, the 1-form $\bZ$ can be gauged away in~\eqref{NUT2} and one obtains static Schwarzschild-Tangherlini-like metrics \cite{Tangherlini63,GibWil87,Birmingham99} of the Robinson-Trautman class~\cite{PodOrt06}, for which the base space can carry any Einstein metric.

\paragraph{Intersection with KS metrics} 

Let us now discuss under what conditions the Einstein spacetimes determined above belong to the KS class, with $\bl_+$ being a (shearfree, twisting) KS vector field. First, since $\bl_+$ is now necessarily an mWAND \cite{OrtPraPra09,MalPra11}, the base space must be K\"ahler-Einstein (as remarked above). Next, by comparing the Riemann component $R_{01ij}$ of~\eqref{R01ij} with the corresponding result for KS spacetimes~\eqref{R01ij_KS} and using~\eqref{NUT_eq} one obtains
\be
 -2\H=\lambda (r^2+\ell^2)+\frac{\mu_0 r}{(r^2+\ell^2)^{\frac{n-2}{2}}} , \qquad \tilde R=n\lambda{\cal F}^2 ,
 \label{NUT_KS_cond}
\ee
where $\mu_0$ is an integration constant. Furthermore, by comparing the component $R_{ijkl}$ of~\eqref{Rijkl} with the KS one~\eqref{Rijkl_KS} one arrives at a constraint on the curvature of the base space (in addition to the already mentioned K\"ahler-Einstein condition), namely
\be
 \tilde R_{\tilde i\tilde j\tilde k\tilde l}=2\lambda(\ell^2\delta_{i[k}\delta_{l]j}+{\cal F}_{\tilde i\tilde j}{\cal F}_{\tilde k\tilde l}-{\cal F}_{\tilde i[\tilde k}{\cal F}_{\tilde l]\tilde j}) .
 \label{NUT_KS_cond2}
\ee
This means that the base manifold is a space of {\em constant holomorphic sectional curvature} \cite{Bochner47,Yanobook_complex,KobNom2}, and therefore it is uniquely given by (assuming it to be simply connected and complete; cf.~Theorems~7.8 and 7.9 of \cite{KobNom2}): (i) the complex projective space ${\mathbb C}P^{\frac{n-2}{2}}$ if $\lambda>0$; (ii) the open unit ball $D^{\frac{n-2}{2}}$ in ${\mathbb C}^{\frac{n-2}{2}}$ if $\lambda<0$; (iii) ${\mathbb C}^{\frac{n-2}{2}}$ (flat space) if $\lambda=0$. This also implies that for $\lambda\neq0$ and $n>4$ it is {\em not} a space of constant curvature.

To summarize, after relabeling $\bl_+=\bk$, the only $n>4$ Taub-NUT vacuum metrics which are also KS are given by the line-element\footnote{To be precise, at this stage we have proven this only in the case $\H C^{-2}\neq$const. However, in section~\ref{subsubsec_NUTspecial} we will prove that the case $\H C^{-2}=$const does not contain any KS-Taub-NUT vacua (except for spacetimes of constant curvature), therefore the general statement about the uniqueness of metric~\eqref{shearfree_seed} is indeed true.}
\be
 \bg=\d r\otimes\bk+\bk\otimes\d r+(r^2+\ell^2)(\bh+\lambda \bk\otimes\bk)+\frac{\mu_0 r}{(r^2+\ell^2)^{\frac{n-2}{2}}}\bk\otimes\bk , \qquad \bk=\d u-2\bZ ,
 \label{shearfree_seed}
\ee
where the base space metric $\bh$ must be K\"ahler-Einstein and of constant holomorphic sectional curvature, with K\"ahler 2-form $\cF=\d\bZ$ (recall~\eqref{F2}, \eqref{NUT_twist}, \eqref{C_canonic}, \eqref{FF_Kahl}, \eqref{Rij_base}, \eqref{NUT_KS_cond2}). Metric~\eqref{shearfree_seed} is indeed of the KS form~\eqref{KS} with $2H=-\mu_0r(r^2+\ell^2)^{\frac{2-n}{2}}$. For $\mu_0\neq0$ the Weyl type is D (cf.~\cite{OrtPraPra13,Ortaggio17,Taghavi-Chabert21}), while $\mu_0=0$ corresponds to a spacetime of constant curvature. For example, for the $n=6$ KS-Taub-NUT metric with $\lambda>0$, the base space is given by ${\mathbb C}P^2$ with the Fubini-Study metric, cf.~\eqref{CP2}, \eqref{CP2_fine}.

The case $n=4$ is special in that~\eqref{NUT_KS_cond2} becomes equivalent to the second of~\eqref{NUT_KS_cond}, which means that the (two-dimensional) base space has constant curvature for any value of $\lambda$. Four-dimensional Taub-NUT metrics of the KS class can thus be written as
\beqn
 & & \bg=\d r\otimes\bl_++\bl_+\otimes\d r+(r^2+\ell^2)\big(2P^{-2}\d\zeta\d\bar\zeta+\lambda\bl_+\otimes\bl_+\big)+\frac{\mu_0 r}{r^2+\ell^2}\bl_+\otimes\bl_+ , 
 \label{NUT_KS_4D} \\ 
 & & \bl_+=\d u-i\ell P^{-1}\big(\bar\zeta\d\zeta-\zeta\d\bar\zeta\big) , \qquad P=1+2\lambda\ell^2\zeta\bar\zeta \qquad (n=4) . \nonumber 
\eeqn
Upon using~\eqref{NUT_u}, metric~\eqref{NUT_KS_4D} corresponds to a fine tuned version of (12.19,\cite{GriPodbook}). For $\mu_0=0$ it reduces to a spacetime of constant curvature.

\subsubsection{Special case $\H C^{-2}=$const}

\label{subsubsec_NUTspecial}

Let us set here $2\H=-c_0 C^2$, where $c_0$ is a constant.\footnote{This special case can also be characterized by $C^2(r)\bl_\pm$ becoming conformal Killing vector fields \cite{Taghavi-Chabert21} (this is true also off-shell, i.e., for any metric~\eqref{NUT} or \eqref{NUT2} with $A^2(r)=2\H(r)=-c_0 C^2(r)$).} The condition $R_{01}=\frac{R}{n}$ with~\eqref{C_canonic} reveals that $c_0$ is fixed by the cosmological constant (recall~\eqref{lambda}) 
\be
  c_0=\lambda .
 \label{NUT_special_R}
\ee

Next, imposing $R_{ij}=\frac{R}{n}\delta_{ij}$ results in 
\be
 \tilde R_{\tilde i\tilde j}=\lambda(2{\cal F}_{\tilde i\tilde k}{\cal F}_{\tilde j\tilde k}+{\cal F}^2\delta_{\tilde i\tilde j}) ,
 \label{NUT_special_ij}
\ee
such that $\tilde R=n\lambda{\cal F}^2$. All components of the Einstein equation are now satisfied.

The final form of the metric is thus given by
\be
 \d s^2=\d r\otimes\bl_+ +\bl_+\otimes\d r+(r^2+\ell^2)(\bh+\lambda\bl_+\otimes\bl_+) , 
 \label{NUT_special}
\ee
where $\bl_+$ and $\ell^2$ are defined in~\eqref{NUT2} and \eqref{C_canonic}, respectively, and the base space metric $\bh$ must obey~\eqref{NUT_special_ij}. It is easy to see (e.g. using~\eqref{NUT_u} backwards) that metric~\eqref{NUT_special} belongs to the class of Brinkmann warps \cite{Brinkmann25} (cf. also, e.g., \cite{petrov,OrtPraPra11} and refs. therein)\footnote{These are the unique Einstein spaces which are properly conformal to other Einstein spaces. The case $n=4$ is special, for such Einstein spacetimes are necessarily of constant curvature \cite{Brinkmann25,petrov,OrtPraPra11}.} --  except when $\lambda=0$.  Let us emphasize that, by~\eqref{NUT_special_ij}, the base metric $\bh$ is not restricted to be Einstein when $\lambda\neq0$, since, generically, ${\cal F}_{\tilde i\tilde k}{\cal F}_{\tilde j\tilde k}$ is not proportional to $\delta_{\tilde i\tilde j}$ (in particular, it cannot be so for an odd $n$). For the same reason, the two null directions defined by~\eqref{NUT_wands} are not WANDs, in general (cf. section~\ref{subsubsec_NUT_curvat}). They are iff $n$ is even and $\bh$ is (almost-)K\"ahler-Einstein, which is the case of the ``Fefferman-Einstein'' metrics studied in \cite{Taghavi-Chabert21}. As in section~\ref{subsubsec_NUT_generic}, they are mWANDs iff, in addition, $\tilde\nabla_{\tilde k}{\cal F}_{\tilde i\tilde j}=0$. The limit $\ell\to0$ gives rise to ``massless'' Schwarzschild-Tangherlini-like metrics with a Ricci-flat base space \cite{Tangherlini63,GibWil87,Birmingham99,PodOrt06}.

Noticing the similarity of~\eqref{NUT_special_ij} to the Einstein equation sourced by a Maxwell field $\cF=\d\bZ$ (in $n-2$ dimensions), it is not difficult to find suitable spatial geometries $\bh$ and thus construct explicit examples of vacuum solutions~\eqref{NUT_special}. For instance, in seven spacetime dimensions with a positive cosmological constant one can take~\eqref{NUT_special} with
\beqn
 & & \bh=\frac{1}{10\lambda\ell^2}\Big[4\left[\d\theta^2+\sin^2\theta(\d\phi^2+\sin^2\phi\d\xi^2)\right]+\d\chi^2+\sin^2\chi\d\psi^2\Big] \qquad (n=7, \lambda>0) , \\
 & & \bZ=\frac{1}{2\sqrt{10}\lambda\ell}\cos\chi\d\psi ,
\eeqn
which is clearly a (non-Einstein) direct product $S^3\times S^2$. Similar solutions can be constructed in other dimensions.

\paragraph{Intersection with KS metrics} 

The first of~\eqref{R01ij} gives
\be
 R_{01ij}=0 .
\ee
Since this contradicts the result~\eqref{R01ij_KS} for KS spacetimes in vacuum, it follows that the two shearfree null directions~\eqref{NUT_wands} cannot be KS vector fields of the spacetime~\eqref{NUT_special} (except in the trivial case $H=0$, i.e., for spacetimes of constant curvature, which is equivalent to~\eqref{shearfree_seed} with $\mu_0=0$).

\section{Integration of the Einstein-Maxwell equations for shearfree twisting KS solutions}
\setcounter{equation}{0}

\label{app_integration}

In this appendix we present the details of the integration of the Einstein-Maxwell equations for the case of a shearfree twisting KS vector field $\bk$, corresponding to the branch of solutions described in section~\ref{subsubsec_shearfree}. As proven in appendix~\ref{app_NUT}, it follows from the results of \cite{Taghavi-Chabert21} that the only $n>4$ KS vacua possessing a shearfree, twisting KS vector field $\bk$ are given by the Taub-NUT metrics~\eqref{shearfree_seed} with~\eqref{NUT_KS_cond2}. In order to charge those spacetimes according to~\eqref{KS}, \eqref{A_KS}, \eqref{H_redef}, we can thus focus on KS line-elements of the form
\be
 \bg=\d r\otimes\bk+\bk\otimes\d r+(r^2+\ell^2)(\bh+\lambda\bk\otimes\bk)-2H(u,r,x)\bk\otimes\bk , \qquad \bk=\d u-2\bZ ,
 \label{shearfree}
\ee
where $\ell$ is a constant. The coordinates $x^\alpha$ ($\alpha=1,\ldots,n-2$, also denoted collectively simply as $x$) parametrize a Riemannian base space of dimension $n-2$ which carries a K\"ahler-Einstein metric $\bh=h_{\alpha\beta}(x)\d x^\alpha\d x^\beta$ of constant holomorphic sectional curvature (cf.~\eqref{NUT_KS_cond2}), and $\bZ=Z_\alpha(x)\d x^\alpha$ is a 1-form which lives in the base space, with K\"ahler 2-form 
given by (cf.~\eqref{F=dZ}, \eqref{F2}, \eqref{C_canonic}, \eqref{FF_Kahl}, \eqref{Rij_base})
\be
 \cF\equiv\d\bZ .
 \label{F=dZ_2}
\ee
The latter is related to the twist matrix of $\bk$ and to the parameter $\ell$ by~\eqref{NUT_twist}, \eqref{FF_Kahl}.

An adapted parallelly transported null frame for metric~\eqref{shearfree} is given by 
\be
  \bk=\pa_r , \qquad \bn=\pa_u+\left[-\textstyle{\frac{\lambda}{2}}(r^2+\ell^2)+H\right]\bk , \qquad \bm_{(i)} = (r^2+\ell^2)^{-\frac{1}{2}}X_i^{\phantom{i}j}(\bmt_{(j)}+2Z_{\tilde j}\pa_u) ,
	\label{frame_KS_NUT}
\ee
where the vectors $\bmt_{(i)} $ define an orthonormal frame of the base space metric $\bh$, the orthogonal matrix $X_i^{\phantom{i}j}$ reads (cf., e.g., appendix~A.5 of \cite{Ortaggio17})
\be
 X_i^{\phantom{i}j}=\ell (r^2+\ell^2)^{-\frac{1}{2}}(\delta_i^j+r\ell^{-2}{\cal F}_{\tilde i}^{\phantom{\tilde i}\tilde j}) ,
\ee
and indices with a tilde denote components of base space quantities in the tilded frame. In the frame~\eqref{frame_KS_NUT}, one finds that the Ricci rotation coefficients needed in the following (along with~\eqref{pt}) are given by\footnote{This can be seen by first obtaining the Ricci rotation coefficients in a simpler but non-parallelly transported frame given by $\bk=\pa_r$, $\bn=\pa_u+\left[-\textstyle{\frac{\lambda}{2}}(r^2+\ell^2)+H\right]\bk$, $\bm_{(i)} = (r^2+\ell^2)^{-\frac{1}{2}}(\bmt_{(i)}+2Z_{\tilde i}\pa_u) $, which follow readily from~\eqref{Li0_KS}--\eqref{L11_KS}, where the background quantities now refer to spacetime~\eqref{shearfree_seed} with $\mu_0=0$ (cf.~\eqref{Lij_NUT}, \eqref{Nij_NUT} with $-2\H=\lambda (r^2+\ell^2)$ and~\eqref{C_canonic}). Using the known transformation rules under spins \cite{OrtPraPra07} one then arrives at~\eqref{Lij_charged}, \eqref{Nij_charged}.}
\beqn
 & & L_{ij}=(r^2+\ell^2)^{-1}\big(r\delta_{ij}+{\cal F}_{\tilde i\tilde j}\big) , \qquad L_{i1}=0=L_{1i} ,  \label{Lij_charged} \\
 & & N_{ij}=\left[-\textstyle{\frac{\lambda}{2}}(r^2+\ell^2)+H\right]L_{ji} , \qquad \qquad  L_{11}=\lambda r-DH . \label{Nij_charged}
\eeqn 
We observe that~\eqref{Lij_charged} gives
\be
 \theta=r(r^2+\ell^2)^{-1} , \qquad A_{ij}=(r^2+\ell^2)^{-1}{\cal F}_{\tilde i\tilde j} ,
 \label{exp_KS_NUT}
\ee
which, with~\eqref{twist}, reveals that the $(a^{0}_{(2 \mu)})^2$ become constants, i.e., 
\be
 (a^0_{(2)})^2=(a^0_{(4)})^2=\ldots=(a^0_{(n-2)})^2=\ell^2 .
 \label{a_shearfree}
\ee 

Further, eq.~\eqref{11k} gives 
\be
 \delta_{j}A_{ji}+\M{k}{j}{j}A_{ki} +\M{k}{i}{j}A_{jk}=0 , 
 \label{11k_red}
\ee 
while \eqref{comm_dD} becomes $\delta_i D-D\delta_i=\theta\delta_i+A_{ji}\delta_j$. The Maxwell component~\eqref{Max3} thus reduces to 
\be
  \left[\delta_iD+2A_{ji}\delta_j +(n-4)\theta\delta_{i}\right]\alpha=0 .
	\label{Max3_KS_NUT}
\ee

Thanks to~\eqref{a_shearfree}, eq.~\eqref{rescale_alpha} with \eqref{beta_even}, \eqref{A_alpha}  takes the form
\be
  \alpha =\frac{r}{(r^2+\ell^2)^{\frac{n-2}{2}}} \Bigg[\alpha_{0}+  \beta_{0}\sum_{\mu=0}^{\frac{n-2}{2}} \binom{\frac{n-2}{2}}{\mu} \frac{\ell^{2\mu}}{n-3-2\mu}r^{n-3-2\mu}\Bigg] .
	\label{alpha_KS_NUT}	
\ee

Using~\eqref{exp_KS_NUT} and \eqref{alpha_KS_NUT}, eq.~\eqref{Max3_KS_NUT} can be written as
\beqn
 (\ell^2-r^2)\delta_i\alpha_{0}+\delta_i\beta_{0}\sum_{\mu=0}^{\frac{n-2}{2}} \binom{\frac{n-2}{2}}{\mu} \frac{(n-4-2\mu)r^2+(n-2-2\mu)\ell^2}{n-3-2\mu}\ell^{2\mu}r^{n-3-2\mu} \nonumber \\
{}+2{\cal F}_{\tilde k\tilde i}\delta_k\Bigg[r\alpha_{0}+  \beta_{0}\sum_{\mu=0}^{\frac{n-2}{2}} \binom{\frac{n-2}{2}}{\mu} \frac{\ell^{2\mu}}{n-3-2\mu}r^{n-2-2\mu}\Bigg]=0 . 
 \label{Max3_KS_NUT_2}
\eeqn

It is easy to see that the highest power of $r$ in the above equation gives $(n-4)\delta_i\beta_{0}=0$, and then (hereafter we assume $n>4$) the subleading terms imply also $\delta_i\alpha_{0}=0$, so that from now on
\be
 \delta_i\beta_{0}=0 , \qquad \delta_i\alpha_{0}=0 .
 \label{alpha_i}
\ee

Next, using \eqref{frame_KS_NUT}, \eqref{Lij_charged}, \eqref{Nij_charged}, \eqref{11k_red} and \eqref{Max1}, the Maxwell component~\eqref{Max2} becomes simply
\be
  \alpha_{,ru}=0 ,
\ee
which means (with~\eqref{alpha_KS_NUT}, \eqref{alpha_i} and \eqref{frame_KS_NUT}) that $\alpha_{0}$ and $\beta_{0}$ are constants, i.e., $\alpha$ is only a function of $r$.

Let us now consider the Einstein equation. Eq.~\eqref{Eij_tracefree} is already satisfied thanks to~\eqref{Lij_charged}, while \eqref{Eii} can be written as
\be
 \frac{r^2}{(r^2+\ell^2)^{\frac{n-4}{2}}}D\left[r^{-1}(r^2+\ell^2)^{\frac{n-2}{2}}2H\right]=-\kappa\left[2\alpha^2\ell^2 +\textstyle{\frac{1}{n-2}}(r^2+\ell^2)^2(D\alpha)^2\right] .
\ee

Since the r.h.s. is a function of $r$ only, this means that 
\beqn
 & & 2H=-r\frac{\mu_0-\kappa f(r)}{(r^2+\ell^2)^{\frac{n-2}{2}}} , \label{H_KS_NUT} \\
 & & \mbox{with} \qquad D\mu_0=0 , \qquad Df=-r^{-2}(r^2+\ell^2)^{\frac{n-4}{2}}\left[2\alpha^2\ell^2 +\textstyle{\frac{1}{n-2}}(r^2+\ell^2)^2(D\alpha)^2\right] . \label{DF}
\eeqn

Now, thanks to~\eqref{Lij_charged} and \eqref{alpha_i} we have (cf.~\eqref{T1i}, \eqref{T11})
\be
 T_{1i}=0 , \qquad T_{11}=0 .
\ee

With~\eqref{R1i} and~\eqref{H_KS_NUT}, the component $(1i)$ of the Einstein equation thus reduces to $\big[(r^2-\ell^2)\delta_i+2r{\cal F}_{\tilde i\tilde j}\delta_j\big]\mu_0=0$, which clearly implies 
\be
 \delta_i\mu_0=0 .
 \label{mu_i}
\ee

Finally, with~\eqref{R11} and \eqref{mu_i}, the component $(11)$ of the Einstein equation becomes simply
\be
  \mu_{0,u}=0 ,
\ee
i.e., $\mu_0$ is a constant.

To summarize, the only charged KS solution admitting an expanding, twisting and shearfree KS vector field is given by metric~\eqref{shearfree} with the vector potential~\eqref{A_KS}, where $\alpha$ and $f$ are functions of $r$ determined by~\eqref{alpha_KS_NUT} and \eqref{DF}, $\mu_0$, $\alpha_0$, $\beta_0$ are integration constants, and $\lambda$ is the cosmological constant (cf.~\eqref{lambda}). Using~\eqref{F_compts}, it follows that the electromagnetic field strength reads
\be
 \bF=(D\alpha)\d r \wedge\bk-2\alpha\cF  , \label{shearfree_F}
\ee
with~\eqref{F=dZ_2} and \eqref{alpha_KS_NUT}. Recall that the K\"ahler-Einstein base space metric $\bh$ must also obey the additional constraint~\eqref{NUT_KS_cond2}, i.e., it is of constant holomorphic sectional curvature.

If desired, one can obtain the explicit form of the function $f(r)$ in~\eqref{H_KS_NUT} by integrating~\eqref{DF}, upon  noticing that (using~\eqref{alpha_KS_NUT} and the binomial theorem)
\be
 D\alpha=\frac{-(n-3)r^2+\ell^2}{r(r^2+\ell^2)}\alpha+\frac{\beta_0}{r} ,
 \label{Dalpha}
\ee
and therefore 
\be
 Df=-\frac{(r^2+\ell^2)^{\frac{n-4}{2}}}{(n-2)r^4}\left\{\alpha^2\big[(n-3)^2r^4+2\ell^2r^2+\ell^4\big]-\beta_0(r^2+\ell^2)^2\left[2\alpha\frac{(n-3)r^2-\ell^2}{r^2+\ell^2}-\beta_0\right]\right\} ,
 \label{DF3}
\ee
with $\alpha$ as in~\eqref{alpha_KS_NUT}.

\providecommand{\href}[2]{#2}\begingroup\raggedright\endgroup

%
%
%
%

\begin{thebibliography}{10}

\bibitem{Kerr63}
R.~P. Kerr, {\it Gravitational field of a spinning mass as an example of
  algebraically special metrics},  {\em Phys. Rev. Lett.} {\bf 11} (1963)
  237--238.

\bibitem{Newmanetal65}
E.~T. Newman, R.~Couch, K.~Chinnapared, A.~Exton, A.~Prakash, and R.~Torrence,
  {\it Metric of a rotating, charged mass},  {\em J. Math. Phys.} {\bf 6}
  (1965) 918--919.

\bibitem{Stephanibook}
H.~Stephani, D.~Kramer, M.~MacCallum, C.~Hoenselaers, and E.~Herlt, {\em Exact
  Solutions of {E}instein's Field Equations}.
\newblock Cambridge University Press, Cambridge, second~ed., 2003.

\bibitem{GriPodbook}
J.~B. Griffiths and J.~Podolsk\'y, {\em Exact Space-Times in {E}instein's
  General Relativity}.
\newblock Cambridge University Press, Cambridge, 2009.

\bibitem{KerSch652}
R.~P. Kerr and A.~Schild, {\it Some algebraically degenerate solutions of
  {E}instein's gravitational field equations},  {\em Proc. Symp. Appl. Math.}
  {\bf 17} (1965) 199--209.

\bibitem{Trautman62}
A.~Trautman, {\it On the propagation of information by waves},  in {\em Recent
  Developments in General Relativity}, pp.~459--463.
\newblock Pergamon Press and PWN, New York and Warszawa, 1962.

\bibitem{DebKerSch69}
G.~C. Debney, R.~P. Kerr, and A.~Schild, {\it Solutions of the {E}instein and
  {E}instein-{M}axwell equations},  {\em J. Math. Phys.} {\bf 10} (1969)
  1842--1854.

\bibitem{Edelen66_I}
D.~G.~B. Edelen, {\it The null-bundle of an {E}instein-{R}iemann space~{I}:
  General theory},  {\em J. Math. Mech.} {\bf 16} (1966) 351--363.

\bibitem{Xanthopoulos83}
B.~C. Xanthopoulos, {\it The optical scalars in {K}err--{S}child-type
  spacetimes},  {\em Ann. Physics} {\bf 149} (1983) 286--295.

\bibitem{ColHilSen01}
B.~Coll, S.~R. Hildebrandt, and J.~M.~M. Senovilla, {\it Kerr-{S}child
  symmetries},  {\em Gen. Rel. Grav.} {\bf 33} (2001) 649--670.

\bibitem{MyePer86}
R.~C. Myers and M.~J. Perry, {\it Black holes in higher dimensional
  space-times},  {\em Ann. Phys. (N.Y.)} {\bf 172} (1986) 304--347.

\bibitem{Chakrabarti86}
A.~Chakrabarti, {\it {K}err metric in eight dimensions},  {\em Phys. Lett. {\rm
  B}} {\bf 172} (1986) 175--179.

\bibitem{EmpRea02prl}
R.~Emparan and H.~S. Reall, {\it A rotating black ring solution in five
  dimensions},  {\em Phys. Rev. Lett.} {\bf 88} (2002) 101101.

\bibitem{PraPra05}
V.~Pravda and A.~Pravdov\'a, {\it {WAND}s of the black ring},  {\em Gen. Rel.
  Grav.} {\bf 37} (2005) 1277--1287.

\bibitem{OrtPraPra09}
M.~Ortaggio, V.~Pravda, and A.~Pravdov\'a, {\it Higher dimensional
  {K}err-{S}child spacetimes},  {\em Class. Quantum Grav.} {\bf 26} (2009)
  025008.

\bibitem{HawHunTay99}
S.~W. Hawking, C.~J. Hunter, and M.~M. Taylor-Robinson, {\it Rotation and the
  {AdS/CFT} correspondence},  {\em Phys. Rev. {\rm D}} {\bf 59} (1999) 064005.

\bibitem{Gibbonsetal05}
G.~W. Gibbons, H.~L{\"u}, D.~N. Page, and C.~N. Pope, {\it The general
  {K}err-de~{S}itter metrics in all dimensions},  {\em J. Geom. Phys.} {\bf 53}
  (2005) 49--73.

\bibitem{Tangherlini63}
F.~R. Tangherlini, {\it Schwarzschild field in $n$ dimensions and the
  dimensionality of space problem},  {\em Il Nuovo Cimento} {\bf 27} (1963)
  636--651.

\bibitem{PodOrt06}
J.~Podolsk\'y and M.~Ortaggio, {\it {R}obinson-{T}rautman spacetimes in higher
  dimensions},  {\em Class. Quantum Grav.} {\bf 23} (2006) 5785--5797.

\bibitem{KunNavPer05}
J.~Kunz, F.~Navarro-L\'erida, and A.~K. Petersen, {\it Five-dimensional charged
  rotating black holes},  {\em Phys. Lett. {\rm B}} {\bf 614} (2005) 104--112.

\bibitem{KunNavVie06}
J.~Kunz, F.~Navarro-L\'erida, and J.~Viebahn, {\it Charged rotating black holes
  in odd dimensions},  {\em Phys. Lett. {\rm B}} {\bf 639} (2006) 362--367.

\bibitem{Frobetal22}
M.~B. Fr{\"o}b, I.~Khavkine, T.~M{\'a}lek, and V.~Pravda, {\it On
  well-posedness and algebraic type of the five-dimensional charged rotating
  black hole with two equal-magnitude angular momenta},  {\em Eur. Phys. J.~C}
  {\bf 82} (2022) 215.

\bibitem{MalPra11}
T.~M\'alek and V.~Pravda, {\it {K}err-{S}child spacetimes with ({A})d{S}
  background},  {\em Class. Quantum Grav.} {\bf 28} (2011) 125011.

\bibitem{OrtPraPra10}
M.~Ortaggio, V.~Pravda, and A.~Pravdov\'a, {\it Type {III} and {N} {E}instein
  spacetimes in higher dimensions: general properties},  {\em Phys. Rev. {\rm
  D}} {\bf 82} (2010) 064043.

\bibitem{OrtPraPra13}
M.~Ortaggio, V.~Pravda, and A.~Pravdov\'a, {\it On the {G}oldberg-{S}achs
  theorem in higher dimensions in the non-twisting case},  {\em Class. Quantum
  Grav.} {\bf 30} (2013) 075016.

\bibitem{OrtPraPra13rev}
M.~Ortaggio, V.~Pravda, and A.~Pravdov\'a, {\it Algebraic classification of
  higher dimensional spacetimes based on null alignment},  {\em Class. Quantum
  Grav.} {\bf 30} (2013) 013001.

\bibitem{Milsonetal05}
R.~Milson, A.~Coley, V.~Pravda, and A.~Pravdov\'a, {\it Alignment and
  algebraically special tensors in {L}orentzian geometry},  {\em Int. J. Geom.
  Meth. Mod. Phys.} {\bf 2} (2005) 41--61.

\bibitem{Coleyetal04}
A.~Coley, R.~Milson, V.~Pravda, and A.~Pravdov\'a, {\it Classification of the
  {W}eyl tensor in higher dimensions},  {\em Class. Quantum Grav.} {\bf 21}
  (2004) L35--L41.

\bibitem{HerOrtWyl13}
S.~Hervik, M.~Ortaggio, and L.~Wylleman, {\it Minimal tensors and purely
  electric or magnetic spacetimes of arbitrary dimension},  {\em Class. Quantum
  Grav.} {\bf 30} (2013) 165014.

\bibitem{GurGur75}
M.~G{\"{u}}rses and F.~G{\"{u}}rsey, {\it {L}orentz covariant treatment of the
  {K}err-{S}child geometry},  {\em J. Math. Phys.} {\bf 16} (1975) 2385--2390.

\bibitem{DerGur86}
T.~Dereli and M.~G{\"{u}}rses, {\it The generalized {K}err-{S}child transform
  in eleven-dimensional supergravity},  {\em Phys. Lett. {\rm B}} {\bf 171}
  (1986) 209--211.

\bibitem{Thompson66}
A.~H. Thompson, {\it A class of related space-times},  {\em Tensor} {\bf 17}
  (1966) 92--95.

\bibitem{Dozmorov70}
I.~M. Dozmorov, {\it Solutions of the {E}instein equations with zero coupling},
   {\em Soviet Physics Journal} {\bf 13} (1970) 1284--1288.

\bibitem{Xanthopoulos78}
B.~C. Xanthopoulos, {\it Exact vacuum solutions of {E}instein's equation from
  linearized solutions},  {\em J. Math. Phys.} {\bf 19} (1978) 1607--1609.

\bibitem{Taub81}
A.~H. Taub, {\it Generalized {K}err--{S}child space-times},  {\em Ann. Phys.}
  {\bf 134} (1981) 326--372.

\bibitem{MonOCoWhi14}
R.~Monteiro, D.~O'Connell, and C.~D. White, {\it Black holes and the double
  copy},  {\em JHEP} {\bf 12} (2014) 056.

\bibitem{ManSte04}
R.~Mann and C.~Stelea, {\it Nuttier {(A)dS} black holes in higher dimensions},
  {\em Class. Quantum Grav.} {\bf 21} (2004) 2937--2961.

\bibitem{Ortaggio17}
M.~Ortaggio, {\it On the uniqueness of the {M}yers-{P}erry spacetime as a type
  {II(D)} solution in six dimensions},  {\em JHEP} {\bf 06} (2017) 042.

\bibitem{Taghavi-Chabert21}
A.~Taghavi-Chabert, {\it Twisting non-shearing congruences of null geodesics,
  almost {CR} structures and einstein metrics in even dimensions},  {\em Ann.
  Mat. Pura Appl.} {\bf 201} (2022) 655--693.

\bibitem{Alekseevskyetal21}
D.~V. Alekseevsky, M.~Ganji, G.~Schmalz, and A.~Spiro, {\it {L}orentzian
  manifolds with shearfree congruences and {K}{\"a}hler-{S}asaki geometry},
  {\em Differ. Geom. Appl.} {\bf 75} (2021) 101724.

\bibitem{OrtPraPra07}
M.~Ortaggio, V.~Pravda, and A.~Pravdov\'a, {\it Ricci identities in higher
  dimensions},  {\em Class. Quantum Grav.} {\bf 24} (2007) 1657--1664.

\bibitem{OrtPodZof08}
M.~Ortaggio, J.~Podolsk\'y, and M.~\v{Z}ofka, {\it {R}obinson-{T}rautman
  spacetimes with an electromagnetic field in higher dimensions},  {\em Class.
  Quantum Grav.} {\bf 25} (2008) 025006.

\bibitem{GibWil87}
G.~W. Gibbons and D.~L. Wiltshire, {\it Space-time as a membrane in higher
  dimensions},  {\em Nucl. Phys. {\rm B}} {\bf 287} (1987) 717--742.

\bibitem{KodIsh04}
H.~Kodama and A.~Ishibashi, {\it Master equations for perturbations of
  generalized static black holes with charge in higher dimensions},  {\em Prog.
  Theor. Phys.} {\bf 111} (2004) 29--73.

\bibitem{KleMorVan98}
D.~Klemm, V.~Moretti, and L.~Vanzo, {\it Rotating topological black holes},
  {\em Phys. Rev. {\rm D}} {\bf 57} (1998) 6127--6137. See also D. Klemm, V.
  Moretti, and L. Vanzo (1999), Erratum: Rotating topological black holes
  [Phys. Rev. D 57, 6127 (1998)], {\em Phys. Rev.} D 60:109902.

\bibitem{CalKle99}
M.~M. Caldarelli and D.~Klemm, {\it Supersymmetry of anti-de~{S}itter black
  holes},  {\em Nucl. Phys. {\rm B}} {\bf 545} (1999) 434--460.

\bibitem{Berard82}
L.~B\'erard~Bergery, {\it Sur de nouvelles vari{\'e}t{\'e}s riemanniennes
  d'{E}instein},  in {\em Institut {E}lie {C}artan}, vol.~6, pp.~1--60.
\newblock Equipe de recherche associ{\'e}e au {CNRS} d'Analyse Globale
  n$^o$~839, {U}niversit{\'e} de Nancy~{I}, Nancy, 1982.

\bibitem{BaiBat85}
F.~A. Bais and P.~Batenburg, {\it A new class of higher dimensional
  {K}aluza-{K}lein monopole and instanton solutions},  {\em Nucl. Phys. {\rm
  B}} {\bf 253} (1985) 162--172.

\bibitem{PagPop87}
D.~N. Page and C.~N. Pope, {\it Inhomogeneous {E}instein metrics on complex
  line bundles},  {\em Class. Quantum Grav.} {\bf 4} (1987) 213--225.

\bibitem{Pravdaetal04}
V.~Pravda, A.~Pravdov\'a, A.~Coley, and R.~Milson, {\it Bianchi identities in
  higher dimensions},  {\em Class. Quantum Grav.} {\bf 21} (2004) 2873--2897.
  See also V. Pravda, A. Pravdov\'a, A. Coley and R. Milson {\em Class. Quantum
  Grav.} {\bf 24} (2007) 1691 (corrigendum).

\bibitem{Coleyetal04vsi}
A.~Coley, R.~Milson, V.~Pravda, and A.~Pravdov\'a, {\it Vanishing scalar
  invariant spacetimes in higher dimensions},  {\em Class. Quantum Grav.} {\bf
  21} (2004) 5519--5542.

\bibitem{ChoPoPSez10}
D.~D.~K. Chow, C.~N. Pope, and E.~Sezgin, {\it Kundt spacetimes as solutions of
  topologically massive gravity},  {\em Class. Quantum Grav.} {\bf 27} (2010)
  105002.

\bibitem{Sokolowskietal93}
L.~M. Sokolowski, F.~Occhionero, M.~Litterio, and L.~Amendola, {\it Classical
  electromagnetic radiation in multidimensional space-times},  {\em Ann.
  Physics} {\bf 225} (1993) 1--47.

\bibitem{OrtPra16}
M.~Ortaggio and V.~Pravda, {\it Electromagnetic fields with vanishing scalar
  invariants},  {\em Class. Quantum Grav.} {\bf 33} (2016) 115010.

\bibitem{Durkeeetal10}
M.~Durkee, V.~Pravda, A.~Pravdov\'a, and H.~S. Reall, {\it Generalization of
  the {G}eroch-{H}eld-{P}enrose formalism to higher dimensions},  {\em Class.
  Quantum Grav.} {\bf 27} (2010) 215010.

\bibitem{Ortaggio14}
M.~Ortaggio, {\it Asymptotic behaviour of {M}axwell fields in higher
  dimensions},  {\em Phys. Rev. {\rm D}} {\bf 90} (2014) 124020.

\bibitem{GolSac62}
J.~N. Goldberg and R.~K. Sachs, {\it A theorem on {P}etrov types},  {\em Acta
  Phys. Polon.} {\bf \textnormal{Suppl.} 22} (1962) 13--23.

\bibitem{KunTho62}
W.~Kundt and A.~Thompson, {\it Le tenseur de {W}eyl et une congruence
  associ\'ee de g\'eod\'esiques isotropes sans distorsion},  {\em C. R. Acad.
  Sc. Paris} {\bf 254} (1962) 4257--4259.

\bibitem{RobSch63}
I.~Robinson and A.~Schild, {\it Generalization of a theorem by {G}oldberg and
  {S}achs},  {\em J. Math. Phys.} {\bf 4} (1963) 484--489.

\bibitem{Lorenz-Petzold87_nut}
D.~Lorenz-Petzold, {\it Higher-dimensional {T}aub-{NUT}-de~{S}itter solutions},
   {\em Prog. Theor. Phys.} {\bf 78} (1987) 11--15.

\bibitem{ChaGib96}
A.~Chamblin and G.~W. Gibbons, {\it Topology and time reversal},  in {\em
  String Gravity and Physics at the {P}lanck Energy Scale} (N.~S\'anchez and
  A.~Zichichi, eds.), vol.~476, pp.~233--253.
\newblock Kluwer Academic Publishers, Dordrecht/Boston/London, 1996.
\newblock \href{http://xxx.lanl.gov/abs/gr-qc/9510006}{{\tt gr-qc/9510006}}.

\bibitem{Taylor-Robinson98}
M.~Taylor-Robinson, {\it Higher dimensional {T}aub-{B}olt solutions and the
  entropy of non compact manifolds},
  \href{http://xxx.lanl.gov/abs/hep-th/9809041}{{\tt hep-th/9809041}}.

\bibitem{AwaCha02}
A.~M. Awad and A.~Chamblin, {\it A bestiary of higher-dimensional
  {T}aub-{NUT}-{AdS} spacetimes},  {\em Class. Quantum Grav.} {\bf 19} (2002)
  2051--2061.

\bibitem{Bochner47}
S.~Bochner, {\it Curvature in {H}ermitian metric},  {\em Bull. Amer. Math.
  Soc.} {\bf 53} (1947) 179--195.

\bibitem{Yanobook_complex}
K.~Yano, {\em Differential Geometry on Complex and Almost Complex Spaces}.
\newblock Pergamon Press, Oxford, 1965.

\bibitem{KobNom2}
S.~Kobayashi and K.~Nomizu, {\em Foundations of Differential Geometry}, vol.~2.
\newblock Interscience, New York, 1969.

\bibitem{AshDas00}
A.~Ashtekar and S.~Das, {\it Asymptotically anti-de~{S}itter spacetimes:
  conserved quantities},  {\em Class. Quantum Grav.} {\bf 17} (2000) L17--L30.

\bibitem{ManSte06}
R.~Mann and C.~Stelea, {\it New {T}aub-{NUT}-{R}eissner-{N}ordstr{\"o}m spaces
  in higher dimensions},  {\em Phys. Lett. {\rm B}} {\bf 632} (2006) 537--542.

\bibitem{Awad06}
A.~M. Awad, {\it Higher dimensional {T}aub-nuts and {T}aub-bolts in
  {E}instein-{M}axwell gravity},  {\em Class. Quantum Grav.} {\bf 23} (2006)
  2849--2859.

\bibitem{DehKoh06}
M.~H. Dehghani and A.~Khodam-Mohammadi, {\it Thermodynamics of
  {T}aub-{NUT}/bolt black holes in {E}instein-{M}axwell gravity},  {\em Phys.
  Rev. {\rm D}} {\bf 73} (2006) 124039.

\bibitem{Brill64}
D.~R. Brill, {\it Electromagnetic fields in a homogeneous, nonisotropic
  universe},  {\em Phys. Rev.} {\bf 133} (1964) 845--848.

\bibitem{Carter68cmp}
B.~Carter, {\it {H}amilton-{J}acobi and {S}chrodinger separable solutions of
  {E}instein's equations},  {\em Commun. Math. Phys.} {\bf 10} (1968) 280--310.

\bibitem{Ruban72}
V.~A. Ruban, {\it Non-singular metrics of {T}aub-{N}ewman-{U}nti-{T}amburino
  type with an electromagnetic field},  {\em Dokl. Akad. Nauk SSSR} {\bf 204}
  (1972) 1086--1089. In Russian.

\bibitem{FloQue19}
D.~Flores-Alfonso and H.~Quevedo, {\it Topological characterization of
  higher-dimensional charged {T}aub-{NUT} instantons},  {\em Int. J. Geom.
  Meth. Mod. Phys.} {\bf 16} (2019) 1950154.

\bibitem{Papapetrou66}
A.~Papapetrou, {\it Champs gravitationnels stationnaires \`a sym\'etrie
  axiale},  {\em Ann. Inst. H. Poincar{\'e} {\em A}} {\bf 4} (1966) 83--105.

\bibitem{Wald74}
R.~M. Wald, {\it Black hole in a uniform magnetic field},  {\em Phys. Rev. {\rm
  D}} {\bf 10} (1974) 1680--1685.

\bibitem{GibPop78}
G.~W. Gibbons and C.~N. Pope, {\it {${\mathbb C}P^2$} as a gravitational
  instanton},  {\em Commun. Math. Phys.} {\bf 61} (1978) 239--248.

\bibitem{PodPap22}
J.~Podolsk\'y and M.~Papaj{\v c}{\'{\i}}k, {\it All solutions of
  {E}instein-{M}axwell equations with a cosmological constant in $2+1$
  dimensions},  {\em Phys. Rev. {\rm D}} {\bf 105} (2022) 064004.

\bibitem{GarciaD17}
A.~A. Garc\'{\i}a~D\'{\i}az, {\em Exact Solutions in Three-Dimensional
  Gravity}.
\newblock Cambridge University Press, 2017.

\bibitem{OrtPodZof15}
M.~Ortaggio, J.~Podolsk\'y, and M.~\v{Z}ofka, {\it Static and radiating
  $p$-form black holes in the higher dimensional {R}obinson-{T}rautman class},
  {\em JHEP} {\bf 1502} (2015) 045.

\bibitem{BahLunWhi17}
N.~Bahjat-Abbas, A.~Luna, and C.~D. White, {\it The {K}err-{S}child double copy
  in curved spacetime},  {\em JHEP} {\bf 12} (2017) 004.

\bibitem{CarPenTor18}
M.~Carrillo-Gonz\'alez, R.~Penco, and M.~Trodden, {\it The classical double
  copy in maximally symmetric spacetimes},  {\em JHEP} {\bf 04} (2018) 028.

\bibitem{Aliev07_2}
A.~Aliev, {\it Electromagnetic properties of {K}err-anti-de~{S}itter black
  holes},  {\em Phys. Rev. {\rm D}} {\bf 75} (2007) 084041.

\bibitem{RobTra62}
I.~Robinson and A.~Trautman, {\it Some spherical gravitational waves in general
  relativity},  {\em Proc. R. Soc. {\rm A}} {\bf 265} (1962) 463--473.

\bibitem{Senovilla15}
J.~M.~M. Senovilla, {\it Black hole formation by incoming electromagnetic
  radiation},  {\em Class. Quantum Grav.} {\bf 32} (2015) 017001.

\bibitem{Carter73}
B.~Carter, {\it Black hole equilibrium states},  in {\em Black holes}
  (C.~De~Witt and B.~S. De~Witt, eds.), pp.~57--214.
\newblock Gordon and Breach, New York, 1973.

\bibitem{Carter68pla}
B.~Carter, {\it A new family of {E}instein spaces},  {\em Phys. Lett. {\rm A}}
  {\bf 26} (1968) 399--400.

\bibitem{Plebanski75_2}
J.~F. Pleba\'nski, {\it A class of solutions of {E}instein-{M}axwell
  equations},  {\em Ann. Physics} {\bf 90} (1975) 196--255.

\bibitem{PleDem76}
J.~F. Pleba\'nski and M.~Demia\'nski, {\it Rotating, charged, and uniformly
  accelerating mass in general relativity},  {\em Ann. Physics} {\bf 98} (1976)
  98--127.

\bibitem{Klemm98}
D.~Klemm, {\it Rotating black branes wrapped on {E}instein spaces},  {\em JHEP}
  {\bf 11} (1998) 019.

\bibitem{Taub51}
A.~H. Taub, {\it Empty space-times admitting a three parameter group of
  motions},  {\em Annals Math.} {\bf 53} (1951) 472--490.

\bibitem{NewTamUnt63}
E.~Newman, L.~Tamburino, and T.~Unti, {\it Empty-space generalization of the
  {S}chwarzschild metric},  {\em J. Math. Phys.} {\bf 4} (1963) 915--923.

\bibitem{PraPraOrt07}
V.~Pravda, A.~Pravdov\'a, and M.~Ortaggio, {\it Type {D} {E}instein spacetimes
  in higher dimensions},  {\em Class. Quantum Grav.} {\bf 24} (2007)
  4407--4428.

\bibitem{Birmingham99}
D.~Birmingham, {\it Topological black holes in anti-de {S}itter space},  {\em
  Class. Quantum Grav.} {\bf 16} (1999) 1197--1205.

\bibitem{Brinkmann25}
H.~W. Brinkmann, {\it Einstein spaces which are mapped conformally on each
  other},  {\em Math. Ann.} {\bf 94} (1925) 119--145.

\bibitem{petrov}
A.~Z. Petrov, {\em Einstein Spaces}.
\newblock Pergamon Press, Oxford, translation of the 1961 {R}ussian~ed., 1969.

\bibitem{OrtPraPra11}
M.~Ortaggio, V.~Pravda, and A.~Pravdov\'a, {\it On higher dimensional
  {E}instein spacetimes with a warped extra dimension},  {\em Class. Quantum
  Grav.} {\bf 28} (2011) 105006.

\end{thebibliography}

\end{document}